\documentclass[aps,
  preprint,
  superscriptaddress,groupedaddress]{revtex4-1}

\usepackage{graphicx}
\usepackage{float}
\usepackage{physics}
\usepackage{amssymb}
\usepackage{color}
\usepackage{tikz}
\usetikzlibrary{shapes}
\definecolor{verde}{RGB}{0,168,89}
\definecolor{rojo}{RGB}{237,50,55}
\definecolor{azul}{RGB}{0,0,255}
\definecolor{violeta}{RGB}{111,0,146}
\definecolor{azulino}{RGB}{0, 33, 116}
\definecolor{black}{RGB}{0,0,0}
\usepackage{xcolor}
\usepackage[T1]{fontenc}
\usepackage{anysize} 
\usepackage{array}
\usepackage{latexsym}
\usepackage[percent]{overpic} 

\newcommand{\verdlin}{\raisebox{2pt}{\tikz{\draw[verde,solid,line width=1pt](0,0) -- (3mm,0);}}}
\newcommand{\verdcirc}{\raisebox{0.5pt}{\tikz{\node[draw,scale=0.5,circle,verde,fill=verde](){};}}}
\newcommand{\blacki}{\raisebox{0.5pt}{\tikz{\node[draw,scale=0.5,circle,black,fill=black](){};}}}

\newcommand{\azcirc}{\raisebox{0.5pt}{\tikz{\node[draw,scale=0.5,regular polygon, regular polygon sides=5,azul,fill=azul](){};}}}
\newcommand{\rojcirc}{\raisebox{0.5pt}{\tikz{\node[draw,scale=0.5,regular polygon, regular polygon sides=4,rojo,fill=rojo](){};}}}
\newcommand{\blackcirc}{\raisebox{0.5pt}{\tikz{\node[draw,scale=0.5,regular polygon, regular polygon sides=3 ,black,fill=black](){};}}}

\newcommand{\blackdotted}{\raisebox{2pt}{\tikz{\draw[azulino,dotted,line width=2.3pt](0,0) -- (6mm,0);}}}
\newcommand{\blackdashdotted}{\raisebox{2pt}{\tikz{\draw[azulino,dashdotted,line width=2.3pt](0,0) -- (6mm,0);}}}

\begin{document}

\title{Dynamic vaccination in partially overlapped multiplex network}

\author{L. G. Alvarez-Zuzek} \email{lgalvere@mdp.edu.ar}
\affiliation{Departamento de F\'{i}sica, Facultad de Ciencias Exactas
  y Naturales, Universidad Nacional de Mar del Plata, and Instituto de
  Investigaciones F\'{\i}sicas de Mar del Plata (IFIMAR-CONICET),
  De\'an Funes 3350, 7600 Mar del Plata, Argentina.}

\author{M. A. Di Muro} \affiliation{Departamento de F\'{i}sica, Facultad de Ciencias Exactas
  y Naturales, Universidad Nacional de Mar del Plata, and Instituto de
  Investigaciones F\'{\i}sicas de Mar del Plata (IFIMAR-CONICET),
  De\'an Funes 3350, 7600 Mar del Plata, Argentina.}

\author{S. Havlin} \affiliation{Department of Physics, Bar-Ilan University, Ramat-Gan 52900, Israel.}

\author{L. A. Braunstein} \affiliation{Departamento de F\'{i}sica,
  Facultad de Ciencias Exactas y Naturales, Universidad Nacional de
  Mar del Plata, and Instituto de Investigaciones F\'{\i}sicas de Mar
  del Plata (IFIMAR-CONICET), De\'an Funes 3350, 7600 Mar del Plata,
  Argentina.}\affiliation{Center for Polymer Studies, Boston
  University, Boston, Massachusetts 02215, USA.}

\date{\today}

\begin{abstract} 

  
In this work we propose and investigate a new strategy of vaccination,
which we call ``dynamic vaccination''. In our model, susceptible
people become aware that one or more of their contacts are infected,
and thereby get vaccinated with probability $\omega$, before having
physical contact with any infected patient. Then, the non-vaccinated
individuals will be infected with probability $\beta$. We apply the
strategy to the SIR epidemic model in a multiplex network composed by
two networks, where a fraction $q$ of the nodes acts in both
networks. We map this model of dynamic vaccination into bond
percolation model, and use the generating functions framework to
predict theoretically the behavior of the relevant magnitudes of the
system at the steady state. We find a perfect agreement between the
solutions of the theoretical equations and the results of stochastic
simulations. In addition, we find an interesting phase diagram in the
plane $\beta-\omega$, which is composed by an epidemic and a
non-epidemic phases, separated by a critical threshold line $\beta_c$,
which depends on $q$. As $q$ decreases, $\beta_c$ increases, $i.e.$,
as the overlap decreases, the system is more disconnected, therefore
more virulent diseases are needed to spread epidemics. Surprisingly we
find that, for all values of $q$, a region in the diagram where the
vaccination is so efficient that, regardless of the virulence of the
disease, it never becomes an epidemic. We compare our strategy with
random immunization and find that using the same amount of vaccines
for both scenarios, we obtain that the spread of the disease is much
lower in the case of dynamic vaccination when compared to random
immunization. Furthermore, we also compare our strategy with targeted
immunization and we find that, depending on $\omega$, dynamic
vaccination will perform significantly better, and in some cases will
stop the disease before it becomes an epidemic.

\end{abstract}

\maketitle
\section{Introduction}

In 2009, the pandemic virus (H1N1) was identified as the cause of many
cases of human illnesses in California and Texas and a severe outbreak
in Mexico \cite{H1N1_fraser,H1N1_meyers,H1N1_r0}. The pandemic had a
reproduction value of approximately $1.5$ and appeared to exhibit a
community transmissibility similar to the respiratory pathogen SARS
coronavirus (SARS-CoV) \cite{riley_03,sars_meyers,Col_07}. Even though
the most commonly affected age group was 5-45 years old, the influenza
also affected other age groups, such as older adults, pregnant women
and children. By the end of the pandemic in 2010, it was registered
that the virus caused the death of around twenty thousand people all
over the world, and it was fueled by the mobility between regions and
different countries. In \cite{Colizza_11} the authors studied the role
of travel restrictions in halting pandemics by using short-range
mobility data, and explored alternative scenarios by assessing the
potential impact of mobility restrictions. However, although this
strategy could be very effective, it was found only useful to slow
down the spread of the diseases, which might give time to the health
authorities to develop a better strategy to stop the epidemic, for
example the development of a new vaccine. Fortunately, in the case of
the H1N1 pandemic, it was possible to develop and deploy a vaccination
campaign, just in time to prevent a further spreading of the disease.

It is well known that infectious diseases usually spread by physical
contact between individuals in a society
\cite{Ander_91,Bailey_75}. Over the years, researchers have found that
the best way to model these types of contact patterns
\cite{Catt_01,gonzalez_08,gardenes_08} is by using the topology of
complex networks \cite{Boc_01,Coh_10, Barrat_04,New_10,past_01}, where
people are represented by nodes, and their interactions, as links. A
commonly-used model for reproducing the dynamics of the spreading of
endemic diseases, such as seasonal influenza or SARS \cite{Col_07}, is
the susceptible-infected-recovered (SIR) model
\cite{New_05,moreno_02,Val_12}. This model groups individuals of a
population into three compartments according to their state:
susceptible (S), infected (I), and recovered (R). When a susceptible
individual is in contact with an infected one, it becomes infected
with probability $\beta$, which is the same for everyone. Infected
individuals recover after a period of time $t_r$, $i.e.$, they become
immunized and cannot be infected again or infect others. When the
parameters $\beta$ and $t_r$ are constant, the effective probability
of infection is given by the transmissibility $T=1-(1-\beta)^{t_r}$
\cite{Dun_01,Coh_hand}. The SIR model has a tree-like structure with
branches of infection that develop and expand, and this is because
infected individuals cannot be re-infected, so the infection can only
move forwards. It has been proven that this process can be mapped into
link percolation \cite{New_03,Bra01}, and thus the dynamic can be
described using the generating function framework. The most important
property in this framework is the probability $f$ that a branch of
infection will expand throughout the network \cite{Bra01}. When a
branch of infection reaches a node with $k$ connections across one of
its links, it can only expand through its $k-1$ remaining
connections. It can be shown that $f$ satisfies the transcendental
equation $f=1-G_1(1-T f)$, where $G_1(x)=\sum_{k=k_{\rm min}}^{k_{\rm
    max}} k P(k)/\langle k \rangle x^{k-1}$, for $x \in [0,1]$, is the
generating function of the underlying branching process
\cite{New_03}. Note that $G_1(1-T f)$ represents the probability that
the branches of infection do not expand throughout the network. In the
steady state of this process, there is a critical threshold $T_c$ that
separates an epidemic phase from a non-epidemic phase. When $T \leq
T_c$ there is an epidemic-free phase with only small outbreaks, which
corresponds to finite clusters in link percolation theory. But, when
$T > T_c$ an epidemic phase develops, the branches of infection
contribute to a spanning cluster of recovered individuals. Thus, the
probability of selecting a random node that belongs to the spanning
cluster is given by $R=1-G_0(1-Tf)$, where $G_0(x)=\sum_{k=k_{\rm
    min}}^{k_{\rm max}}P(k)x^k$ is the generating function of the
degree distribution.

The spread of epidemics in networks
\cite{castellano_10,Pastor_15,castellano_10,Arenas_16,Braunstein_16}
have been the focus of motivation of several investigations that seek
to develop and study different strategies of mitigation for decreasing
the impact of diseases on healthy populations
\cite{New_05,Buo_13,Granell_13_1,Cozzo_13,Sanz_14,Lag_01}. On one
hand, referring to non-pharmaceutical strategies, one of the most
common and studied is ``quarantine'', in which all individuals of the
affected population must remain in isolation for a period of
time. This scenario is difficult to perform, besides it involves a
great economic loss. On the other hand, a more moderate strategy
proposed is ``social distancing''
\cite{wang_coupled,Valdez_13,wang_12}. In this strategy, susceptible
individuals distance themselves from infected or from those having the
symptoms of the disease, by removing links to them. Although all these
strategies are beneficial, without any doubt, the most effective one
is ``vaccination'' \cite{vaccination16} (and references therein). In
the early times {\it random vaccination} has been studied
\cite{past_03}. In this protocol susceptible individuals are
vaccinated regardless of whether there are or not in contact with an
infected individual, which requires a huge amount of vaccines and
resources that may not be available. One way to improve this strategy
is {\it targeted vaccination} \cite{madar_04}, that is to vaccinate
those people that have many connections, and therefore, higher
probability of getting the disease and transmitting it. However, this
strategy is difficult to implement, since the degrees of nodes in the
network must be known in detail. In \cite{Coh_03}, the authors
proposed the {\it acquaintance immunization} strategy which does not
require knowledge of degrees. In the model, a fraction of nodes is
choosen at random, then some of their closest contacts get
immunized. So, given that in scale-free networks a randomly chosen link
points with high probability to a high degree node, this strategy is
actually a preferential immunization of the hubs, which reduces
dramatically the amount of vaccines needed to control the epidemic.

Since the advent of multilayer networks or network of networks
these structures have been the focus of much research and have allowed
the scientific community to use a more realistic approach. With this
new insight, the spread of epidemics was once again a thrive subject
to investigate
\cite{Dickison_12,Men_12,Yag_13,Sanz_14,Buono_14,arruda_17}. On top of
that, the cases of the H1N1 pandemic (2009) and the Ebola outbreak in
Africa (2014), and their catastrophic consequences worldwide, have
prompted the research of new mitigation strategies to avoid similar
damages in future epidemics outbreaks
\cite{tizzoni2012real,Valdez_Ebola,Gomes_14,wang_12,Alvarez-zuzek_15,wang_acquaintance}.
With this motivation in mind, we develop a vaccination strategy, which
we called ``dynamic vaccination'' on the topology of a multiplex
network. We are interested in studying how epidemic spreads in the
presence of this new protocol of vaccination. In our model, the group
of susceptible people that have a relation with infected individuals,
is the target of the immunization strategy. The contacts or neighbors
of an infected person receives a vaccine with a probability
$\omega$. If the susceptible individuals are vaccinated, they acquire
immunization and cannot be infected or infect others anymore.

This strategy, known also in the epidemiology field as ``ring
vaccination'', has been applied to eradicate the smallpox
\cite{Smallpox} and has been studied in \cite{perisic_09,muller_ring}
due to its efficiency as a vaccination protocol. It has been also
implemented during the Ebola virus epidemic in West Africa (2015)
\cite{merler_16,lancet_ebola}. In this manuscript we present new
insights by applying the strategy on the SIR epidemic model with the
topology of a multiplex network. We perform stochastic simulations and
we apply the generating functions framework to develop theoretical
equations that describe the outcome of the model. We find an excellent
agreement between the simulation results and the solutions of the
analytical equations.

This manuscript is organized as follows: Section II explains the model
and the results obtained, and this section is divided in three parts
$A$, $B$ and $C$. In $A$ we present the rules of the epidemic model
with the dynamic vaccination. Then, in $B$ we develop the theory that
corresponds to the epidemic spreading problem. Lastly, in $C$ we show
the simulations results and compare them with the results of the
theoretical equations, which are solved numerically. Finally, in
Section III we present our discussions and outlooks.

\section{Model and Results}

Our epidemic model is performed on a multiplex system composed of two
networks $A$ and $B$, each one characterized by a degree distribution
$P^\alpha(k)$, with $\alpha=A,B$. Both networks are connected to each
other through a fraction $q$ identical pairs of nodes in both networks
\cite{Buono_14}. Those pairs of nodes in network $A$ and $B$ represent
the same individual acting in different networks. For example one
network could represent the personal contacts of the individuals at
their workplace, and the other network their family or friends.

\subsection{Model}

\begin{figure}[!t]
  \begin{center}
    \includegraphics[width=0.85\linewidth]{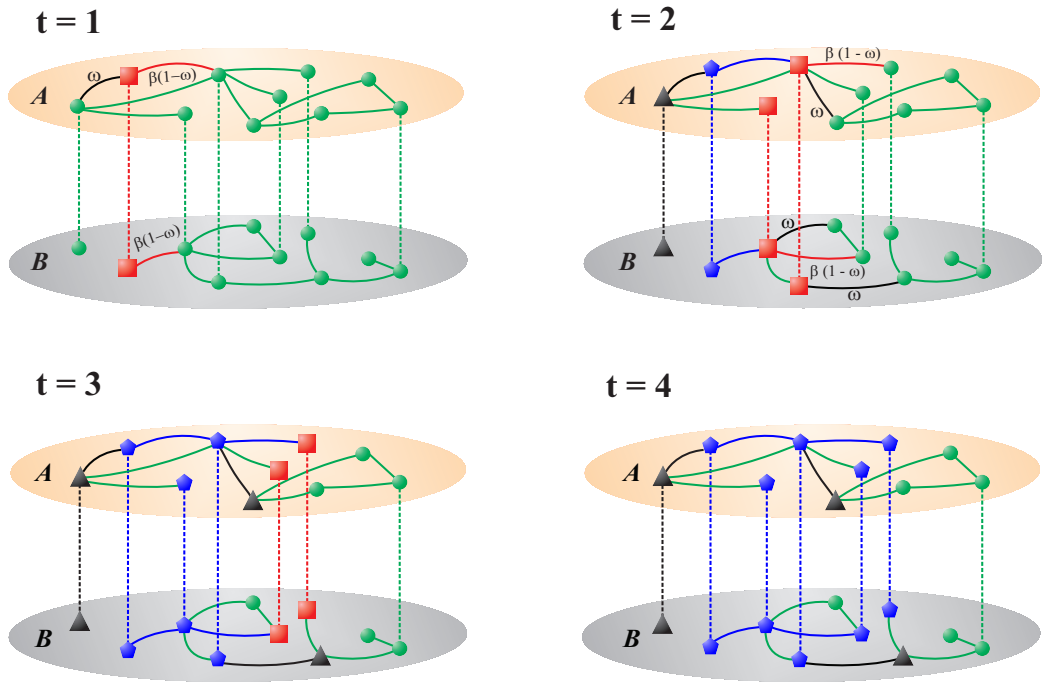}
  \end{center}
  \caption{{\bf Schematic of the SI-R/V epidemic process} in a
    multiplex network consisting of two networks, each of size
    $N=10$. A fraction $q$ of nodes in both networks represent the
    same individual acting in different environments. Here
    $q=0.7$. The colors of the nodes represent the following states:
    green (\protect\verdcirc) for Susceptible (S), red
    (\protect\rojcirc) for Infected (I), blue (\protect\azcirc) for
    recovered (R) and black (\protect\blackcirc) for vaccinated
    individuals. In this case, we assume $t_r = 1$. At $t=1$, the
    patient zero infects one of its neighbors in network $A$ and the
    other becomes vaccinated. Since this infected patient is present
    in both networks, it can also spread the disease in network
    $B$. The red lines indicate the direction of the branches of
    infection. In the next time step, $t=2$, this patient zero recover
    in both networks and the new infected individuals continue
    spreading the disease. The process continues until the fourth
    temporary step, which is the steady state, where there are no more
    infected individuals that can continue spreading the disease.}
  \label{Esquematico}
\end{figure}

At the initial stage of the Susceptible-Infected-Vaccinated-Recovered
model (SI-R/V) all individuals in both networks are susceptible. We
randomly infect an individual in network $A$, which we call patient
zero, and also its counterpart in network $B$, in case it belongs to
both networks. Then, all the neighbors of this patient zero (in both
networks, $A$ and $B$) will be vaccinated and hence immunized with
probability $\omega$. For the sake of simplicity, we consider that
this probability is the same for all individuals. On the other hand,
those neighbors who did not get vaccinated will be infected with
probability $(1-\omega) \beta$. Once an individual receive a vaccine,
it can no longer acquire the disease and becomes immunized. Infected
individuals will recover after $t_r$ time steps and can not be
affected by the disease anymore. In Fig.~\ref{Esquematico} we
illustrate the dynamic of our model for $t_r=1$ and $q=0.7$.

We illustrate the model in Fig.~\ref{Esquematico}. At the beginning of
the dynamic, at $t=1$, there is only one infected node (patient zero -
in red). In network $A$, this individual has two neighbors: one of
them gets vaccinated with probability $\omega$, and the other one
becomes infected with probability $(1-\omega)\beta$. As both
individuals are present in both networks, their counterparts in
network $B$ are also vaccinated and infected, respectively. At $t=2$,
patient zero recovers in both networks and the new infected
individuals will try to infect their susceptible neighbors. In network
$A$, one of the new infected individuals has only one vaccinated
neighbor, thus he cannot spread the disease, and the other has three
susceptible neighbors: one gets vaccinated, the other becomes
infected, as well as his counterpart in network $B$, while the last
remains susceptible. In network $B$ there are two infected
individuals, of which only one manages to infect one of its
neighbors. At the next time step, this infected individual recovers
and the newly infected individuals continue to spread the disease. The
dynamic continues until the system reaches a steady state in which
there are no more infected nodes ($t=4$) and the epidemic process
ends.

\subsection{Theory}
\label{theory}

In our model of the SI-R/V process we assume that the transmissibility
is the same in both networks, thus, all the individuals in the system
spread the infection with the same probability. As we mentioned
earlier, each infected individual in each network can infect each one
of its neighbors (with probability $\beta$), if they have not been
vaccinated earlier(with probability $(1-\omega)$). Thus, at each time
step the probability that an infected node infects a susceptible
neighbor is $(1-\omega) \beta$ during a period of time $t_r$, after
which he recovers. Then, the overall transmissibility $T_{\beta}\equiv
T(\beta,t_r,w)$ is the probability that an infected individual will
transmit the disease to its neighbors, which is given by,
\begin{equation}
  T_{\beta} = (1-\omega) \beta
  \sum_{t=1}^{t_r}[(1-\omega)(1-\beta)]^{t-1} =
  \frac{1-(1-\omega)^{t_r}(1-\beta)^{t_r}}{\omega+\beta-\omega \;
    \beta} \; (1-\omega) \; \beta \;.
  \label{transmissibility}
\end{equation}

We can use the generating function framework to map this process onto
link percolation in a system of two coupled networks
\cite{New_03,Bra01}, after which we can write two transcendental
coupled equations for $f_{A/B}$. Thus, the probability of reaching,
through a random chosen edge, a node that belongs to a branch of
infection that expands all over the system, is,
%
\begin{widetext}
  \begin{align}
    f_A  &= (1-q)\; [1-G_1^A(1-T_{\beta}f_A)]+q \;[1-G_1^A(1-T_{\beta}f_A)\;G_{0}^B(1-T_{\beta}f_B)], \; \nonumber\\
    f_B  &= (1-q)\; [1-G_1^B(1-T_{\beta}f_B)]+q \;[1-G_1^B(1-T_{\beta}f_B)\;G_{0}^A(1-T_{\beta}f_A)]. \; 
    \label{rama}
  \end{align}
\end{widetext}
Here, $G_0^{A/B}(x)$ and $G_1^{A/B}(x)$ are the generating functions
of the degree and the excess degree distributions for each network
respectively. On one hand, $G_1^{A/B}(x)$ represents the probability
that choosing a random edge that leads to a node of degree $k$ in one
network, this branch cannot spread the disease through its remaining
$k-1$ connections. On the other hand, $G_0^{A/B}(x)$ takes into
account the probability that, if the node has a counterpart node of
degree $k$ in the other layer ($B/A$), the branch of infection does
not spread through its $k$ links, $i.e.$, the branch does not spread
in layer $B/A$.

In Eqs. (\ref{rama}) the first term in both equations corresponds to
those branches of infection that only spread within their own network,
while the second term takes into account those branches that spread
through both networks.

During the dynamics, the branches of infection reaches both recovered
and vaccinated nodes. The difference is that once the infection branch
crosses a link to reach a node that has been vaccinated, this
vaccinated individual cannot spread the disease. Thus, we can develop
in the same way as before, a transmissibility $T_{\omega}\equiv
T(\beta,t_r,w)$ as the effective probability that a susceptible
neighbor in contact with an infected node, for a period of time $t_r$,
will be vaccinated. This transmissibility is given by,
\begin{equation}
  T_{\omega} =\omega \sum_{t=1}^{t_r}  [(1-\omega)(1-\beta)]^{t-1} =
  \frac{1-(1-\omega)^{t_r}(1-\beta)^{t_r}}{\omega+\beta-\omega \;
    \beta} \; \omega .
  \label{transmissibility2}
\end{equation}
Therefore, in the steady state of our model, the magnitude that maps
with the order parameter of link percolation is not only the fraction
of recovered individuals, as in the standard SIR \cite{New_05}, but
instead, it is the sum of vaccinated and recovered, i.e.,
V~$+$~R. This is due to the fact that the infection branches also
reach those nodes that were vaccinated via a link. A susceptible
individual can get vaccinated only if he has an infected neighbor that
could infect him through a S-I link. To consider this event in the
equations, a multiplicative factor must be added, which takes into
account the probability that a susceptible node becomes infected or
vaccinated if its state is altered by one of its infected neighbors.

Then, for the parameter $R$, we should consider the probability that a
randomly chosen node is connected to a branch of infection through at
least one of it's $k$ links. Thus, the fraction of recovered
individuals in each network can be written as,

\begin{align}
  R_A &= \frac{T_{\beta}}{T_{\beta}+T_{\omega}} \; \{ (1-q)\; [1-G_0^A(1-T_{\beta}f_A)]+q\;[1-G_0^A(1-T_{\beta}f_A)\;G_{0}^B(1-T_{\beta}f_B)]\} \; , \nonumber\\
  R_B &= \frac{T_{\beta}}{T_{\beta}+T_{\omega}} \{(1-q)\; [1-G_0^B(1-T_{\beta}f_B)]+q\;[1-G_0^B(1-T_{\beta}f_B)\;G_{0}^A(1-T_{\beta}f_A)] \} \; .
  \label{R}
\end{align}
And the fraction of vaccinated nodes is given by,
\begin{align}
  V_A &= \frac{T_{\omega}}{T_{\omega}+T_{\beta}} \; \{ (1-q)\; [1-G_0^A(1-T_{\beta}f_A)]+q\;[1-G_0^A(1-T_{\beta}f_A)\;G_{0}^B(1-T_{\beta}f_B)]\} \; , \nonumber\\
  V_B &= \frac{T_{\omega}}{T_{\omega}+T_{\beta}} \{(1-q)\; [1-G_0^B(1-T_{\beta}f_B)]+q\;[1-G_0^B(1-T_{\beta}f_B)\;G_{0}^A(1-T_{\beta}f_A)] \} \; .
  \label{V}
\end{align}
Then, the total fraction of recovered (R) and vaccinated (V)
individuals in the system is given by,
\begin{align}
  R &= (R_A+R_B-\zeta_R)/(2-q) \nonumber\\
  V &= (V_A+V_B-\zeta_V)/(2-q),
  \label{RT}
\end{align}
where $\zeta_R = \frac{T_{\beta}}{T_{\omega}+T_{\beta}} q
[1-G_0^A(1-T_{\beta}f_A) G_0^B(1-T_{\beta}f_B)]$ is the fraction of
shared nodes that are recovered and $\zeta_V =
\frac{T_{\omega}}{T_{\beta}+T_{\omega}} q [1-G_0^A(1-T_{\beta}f_A)
  G_0^B(1-T_{\beta}f_B)]$ is the fraction of shared nodes that are
vaccinated, in the steady state.

From Eqs. (\ref{transmissibility}) and (\ref{rama}) we can see that if
we use the total transmissibility, which is the sum of $T_{\beta}$ and
$T_{\omega}$, as the control parameter we lose information about the
probability of vaccination, $\omega$ (see Appendix
\ref{appen1}). Hence, we will make use of the virulence of the
diseases $\beta$ as the control parameter. To this end, we
fixed $t_r=1$ and obtain $\beta$ by inverting
Eq. (\ref{transmissibility}).

\subsection{Simulation Results}

In the simulations, we generate two uncorrelated networks, $A$ and
$B$, of equal size using the Molloy-Reed algorithm \cite{Mol_01}. We
randomly overlap a fraction $q$ of nodes in network $A$ with nodes in
network $B$ by a one-to-one connection. The degree distribution in
each network is given by $P_i(k)$, with $i=A,B$ and $k_{\rm min} \leq
k \leq k_{\rm max}$, where $k_{\rm min}$ and $k_{\rm max}$ are the
minimum and the maximum degree that a node can have.  We assume that
an epidemic takes place in a single realization if the number of
recovered individuals is larger than a certain value $s_c$. Since we
consider networks of size $N=10^5$ we chose $s_c=200$
\cite{Lag_02,Yanqing15}. To calculate the total number of recovered
and vaccinated nodes throughout the entire system, the pair of nodes
that act in both networks are counted as single nodes. For the sake of
simplicity we set $t_r=1$.

\begin{figure}[t!]
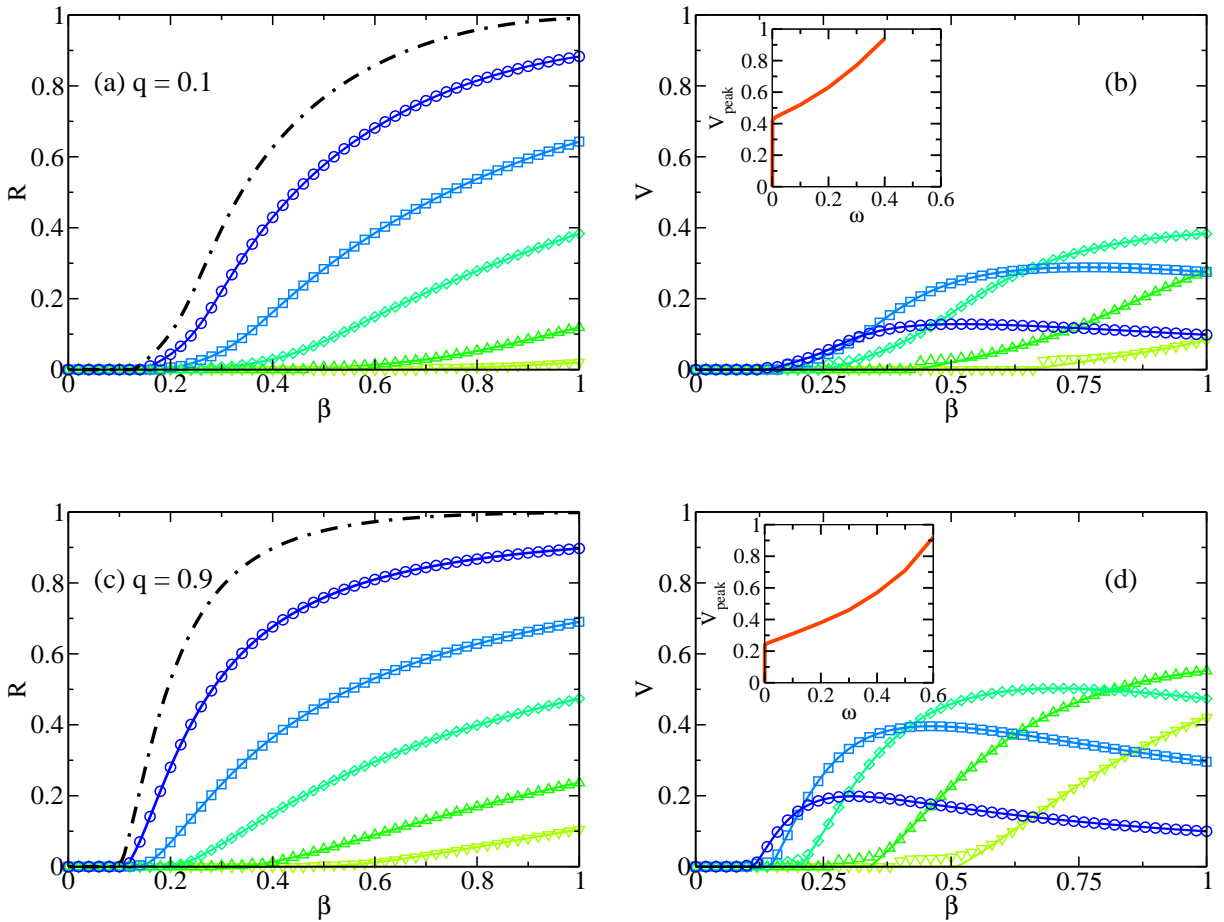

  \centering
  \begin{minipage}[c]{\textwidth}
    \centering
    \includegraphics[width=3.0in]{2a.eps}\hspace{0.5cm}
    \includegraphics[width=3.0in]{2b.eps}\vspace{1cm}
    \includegraphics[width=3.0in]{2c.eps}\hspace{0.5cm}
    \includegraphics[width=3.0in]{2d.eps}
    \caption{{\bf Theoretical and Simulation results} of the total
      fraction of recovered (R) and vaccinated (V) nodes, for two
      different overlap scenarios ($q=0.1,0.9$), as a function of the
      virulence of the diseases $\beta$ and for different values of
      $\omega$, in the steady state of the process. We consider
      $t_r=1$, (a)-(b) $q=0.1$ and (c)-(d) $q=0.9$, and
      $\omega=0,0.1,0.3,0.5,0.7,0.8$ from left to right. The symbols
      correspond to the simulation results and the lines correspond to
      the theoretical evaluation of Eqs.~(\ref{RT}). The multiplex
      network consists of two layers, each of size $N= 10^5$. Layer $A$
      is a ER network with $\langle k_A \rangle = 4$ with $k_{min}=0$
      and $k_{max}= 40$, and layer $B$ is a SF network with $\lambda_B
      = 2.5$, $k_{min}=2$, $k_{max}=\sqrt {10^5}$ and exponential cutoff $c
      = 50$, thus $\langle k_B \rangle \simeq 3.66$. Simulation
      results are averaged over $10^4$ realizations.}
    \label{Fig1}
  \end{minipage}
\end{figure}

In Fig.~\ref{Fig1} we show the total number of recovered (R) and
vaccinated (V) nodes as a function of $\beta$ for two different
vaccination scenarios, and we vary the overlap $q$ between the
layers. To see the effect of the multilayer structure we consider two
different networks: layer $A$ is an Erd\H{o}s-R\'enyi (ER) network
\cite{Erd_01}, with average degree $\langle k_A \rangle = 4$,
$k_{min}=0$ and $k_{max}=40$. Nodes with a connectivity equal to
$k_A=0$ are isolated in layer $A$ and can only be infected in layer
$B$. Thus, this nodes do not play a major role in the spreading
process between layers.  The layer $B$ is a truncated scale free
network (SF) where $P_B(k_i) \sim k_i^{-\lambda_B} e^{-k_i/c}$ with
$\lambda_B=2.5$, $k_{min}=2$, $k_{max}=\sqrt {10^5}$ and an
exponential cutoff $c=50$ \cite{New_03}. Both networks are with
$N=10^5$ nodes. Each curve corresponds to a different value of
$\omega=0,0.1,0.3,0.5,0.7,0.8$ from left to right. We consider the
cases $q=0.1$ ((a) and (b)), and $q=0.9$ ((c) and (d)). Insets in
Fig.~\ref{Fig1} (b) and \ref{Fig1}(d) correspond to the maximum in the
fraction of vaccinated individuals $V_{peak}$, if there is one, as a
function of $\omega$.

Figure \ref{Fig1} shows excellent agreement between simulation results
and the theoretical analysis (Eqs. (\ref{R}) and(\ref{RT})). As
expected, the critical threshold $\beta_c$ increases as $q$
decreases. Hence, for less interconnected networks a more virulent
disease is needed in order to become an epidemic. For instance, for
$\omega=0.8$ there is a noticeable difference in $\beta_c$, when $q$
is low the reach of the disease is insignificant while for $q$ closed
to one reaches $10\%$ of the healthy population. On top of that, it
can be seen that qualitatively, the behaviors of both magnitudes, are
very similar regardless the overlap.

In Fig.~\ref{Fig1}(a) and \ref{Fig1}(c) we can see that as $\omega$
increases the total fraction of recovered nodes decreases. For high
values of $\beta$, such as $\beta=1$, when $\omega=0.1$ the disease
reaches $90\%$ of the population, regardless of the overlap. But as
$\omega$ increases, the immunized individuals block many of the paths
that would be used by the disease to spread through the
population. This causes a decrease in the probability that the disease
might spread or ``percolate'' through an edge. Thus, as the
probability of vaccination gets higher, the disease has to be more
virulent to reach the entire system, which translates into an increase
in the critical infection threshold $\beta_c$. Notice that for $\omega
\gtrsim 0.8$, the disease never originates an epidemic despite its
virulence.

In contrast, in Fig.~\ref{Fig1}(b) and \ref{Fig1}(d), we can see that
the total fraction of vaccinated nodes does not behave monotonically
with $\beta$. For small $\beta$, V increases until it reaches a
maximum value, then starts to decrease. This maximum value varies with
$\omega$, is more pronounced for low values of $\omega$ and vanishes
as $\omega$ becomes larger. The inset in each plot shows this peak as
a function of $\omega$. When the probability of vaccination is equal
to zero, there is no vaccinated nodes. However, slightly above
$\omega=0$ a peak exhibits. Then, as $\omega$ increases the fraction
of vaccinated individuals in the peak also increases reaching a
certain value of $\omega$ above which the peak vanishes. Note that
this maximum value also varies with the overlap fraction, $q$.

The origin of the peaks is the competition between the spread of the
disease and the vaccination process. For instance, let's consider the
case of low values of $\omega$, such as $\omega = 0.1$. Slightly above
$\beta_c$, the disease is not virulent enough so, increasing $\beta$
the number of infected and vaccinated nodes increases. This is true
until a certain value of $\beta$, that corresponds to the peak in V,
above which individuals are more likely to become infected rather than
vaccinated. On the contrary, in the case of high $\omega$ values, the
peak is not exhibited. This is due to the fact that the virulence of
the disease is not strong enough to overcome the
vaccination. Furthermore, as $\omega$ increases and for high values of
overlap, the fraction of vaccinated individuals is much higher than
for low values of $q$.

\begin{figure}[t!]
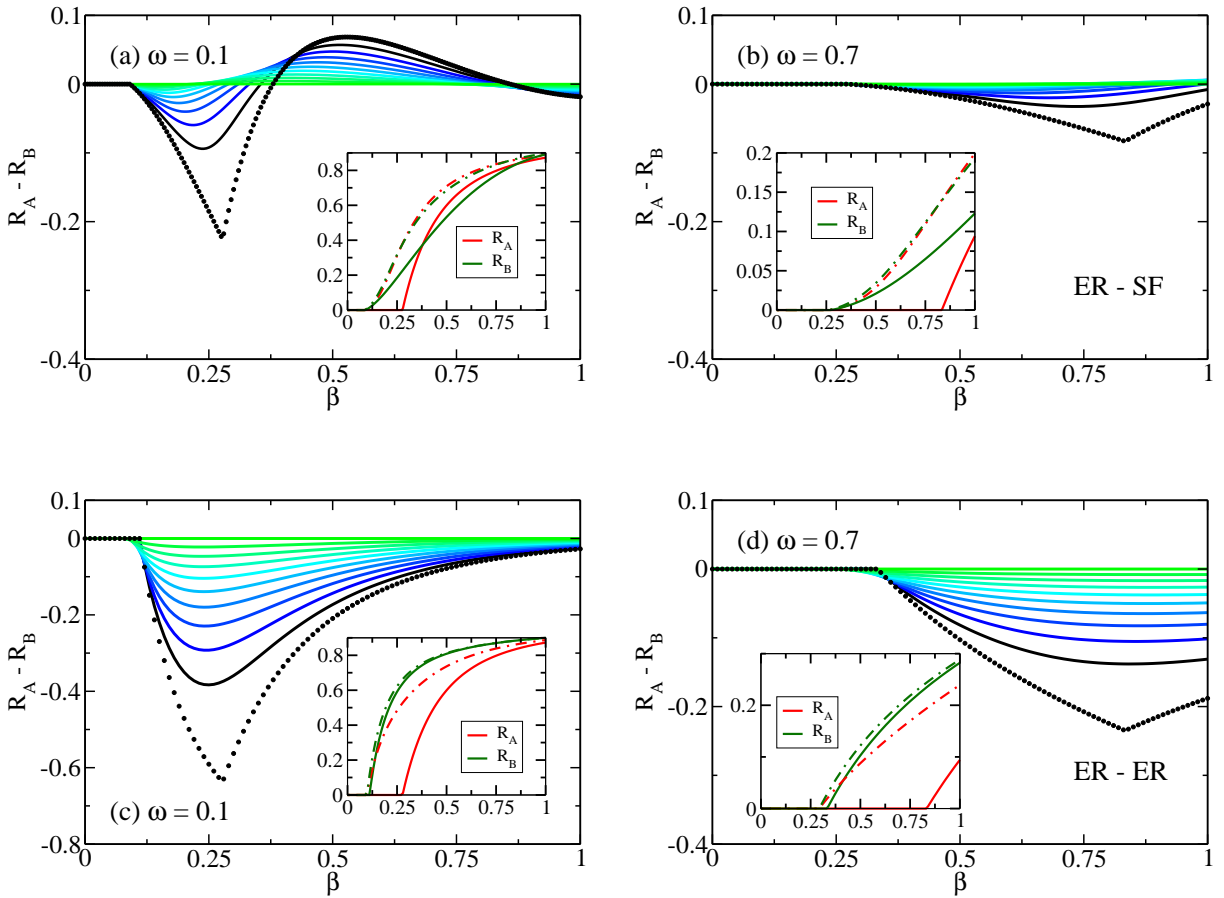

  \centering
  \begin{minipage}[c]{\textwidth}
    \centering
    \includegraphics[width=3.0in]{3a.eps}\hspace{0.5cm}
    \includegraphics[width=3.0in]{3b.eps}\vspace{1cm}
    \includegraphics[width=3.0in]{3c.eps}\hspace{0.5cm}
    \includegraphics[width=3.0in]{3d.eps}
    \caption{{\bf Difference between the fraction of recovered and
        vaccinated individuals in both layers} as a function of the
      virulence of the disease for different values of the overlap
      $q$. We consider $\omega=0.1$(left)$,0.7$(right) and
      $t_r=1$. The overlap $q$ varies from $0$ (\protect\blacki) to
      $1$ (\protect\verdlin) with $\Delta q = 0.1$. The curves
      correspond to the theoretical evaluation of Eqs.~(\ref{RT}). In
      (a) and (b) we consider in layer $A$ an ER network with average
      degree $\langle k_A \rangle = 4$, $k_{min}=0$ and $k_{max}=40$,
      and in layer $B$ a SF network with $\lambda_A=\lambda_B=2.5$,
      $k_{min}=2$, $k_{max}=\sqrt {10^5}$, and exponential cutoff
      $c=135$, thus $\langle k_B \rangle \simeq 4$. Both networks are
      of the same size $N=10^5$. In (c)-(d) the system is composed by
      two ER, layer $A$ with $\langle k_A \rangle = 4$ and layer $B$
      with $\langle k_B \rangle = 10$, both with $k_{min}=0$ and
      $k_{max}= 40$. The insets show the fraction of recovered
      individuals in each layer for the case of $q=0$ (solid lines)
      and $q=0.5$ (dash-dotted lines).}
   \label{Fig3}
  \end{minipage}
\end{figure}

Now, in order to study the interplay between the layers we vary the
overlap in the system and consider a scenario of low and high
probability of vaccination, $\omega = 0.1$ (left) and $\omega=0.7$
(right). We show in Fig.~\ref{Fig3} the difference between the total
number of recovered individuals ($R_A$ - $R_B$) of each layer as a
function of $\beta$. Each curve corresponds to different values of
$q$, with $q \in [0,1]$, and $\Delta q = 0.1$.  In
Fig.~\ref{Fig3}(a)-(b) we consider in layer $A$ an ER network with
average degree $\langle k_A \rangle = 4$, $k_{min}=0$ and
$k_{max}=40$, and in layer $B$ a SF network with
$\lambda_A=\lambda_B=2.5$, $k_{min}=2$, $k_{max}=\sqrt {10^5}$ and an
exponential cutoff $c=135$, thus $\langle k_B \rangle \simeq 4$
\cite{New_03}. Note that this type of SF networks is appropriate to
describe scenarios and structures of real-word systems \cite{Ama_01}.
For the case of isolated networks, $q=0$, we can observe that there is
a notable difference between the fraction of recovered individuals in
each layer. For low values of $\beta$ it is well known that the level
of the epidemic is higher in layer $B$ due to the ultra small world
property \cite{cohen_ultrasmall}. In Fig.~\ref{Fig3}(a), when the
probability of vaccination is low, there are two regimes: for low
values of $\beta$, the disease spreads more in the SF network, but
above a certain $\beta$ value the ER network topology is more
efficient to propagate the infection. In Fig.~\ref{Fig3}(c), the
networks have the same topology, and thus, it is expected that there
will be more recovered individuals in the network (layer $B$) with
larger average degree.

In Fig.~\ref{Fig3}(b) we can observe that for $\omega=0.7$ the
vaccination strategy is very effective, and the difference between the
spreading of the disease is low. As we can see from the inset, this is
due to the fact that the disease almost does not spread. Besides,
when $q=0$ (solid lines), the critical value above which the epidemic
occurs, is different for each network and much lower for the SF
network. But for the case of $q=0.5$ (dash-dotted lines) this value is
dominated by the network where the epidemic spreads more
easily. Besides, for $\beta=1$ the infection reaches nearly a $10\%$
of the population, and even more in the SF network. However, when the
networks are overlapped the infection spreads more easily, and hence
as $q$ increases the difference between the $R_A$ and $R_B$ shrinks to
zero.

Focusing on the critical threshold $\beta_c$, this can be obtained
theoretically from the intersection of the two Eqs. (\ref{rama}) where
all branches of infection stop spreading, which is given when $f_A =
f_B = 0$. This is equivalent to find the solution of the system
$det(J-I)=0$ where I is the identity matrix and J is the Jacobian
matrix of the coupled equation with $J_{i,k}|_{f_i=f_k=0}=\partial f_i
/ \partial f_k|_{f_i=f_k=0}$. In \cite{Buono_14}, this theoretical
critical value has been obtained for standard SIR model in a partially
overlapped multiplex networks, $T_c^{SIR}$, and it is equal to the
transmissibility of our model, $T_c^{SIR}=T_c$. However, the critical
infection probabilities are different in both models. For instance,
for $t_r=1$ it is well known that $T_c^{SIR}=\beta_c^{SIR}$, and in
our model $T_c=\beta_c(1-\omega)$. Thus
$\beta_c^{SIR}=\beta_c(1-\omega)$. Next, to account how the magnitudes
change with $\omega$, we will focus on the critical value of virulence
of the diseases, $\beta_c=\beta_c^{SIR}/(1-\omega)$ and not on the
critical transmissibility. Using numerical evaluations we find a
physical and stable solution for $\beta_c$.

\begin{figure}[t!]
  \centering
  \begin{minipage}[c]{\textwidth}
    \centering
    \includegraphics[width=3.0in]{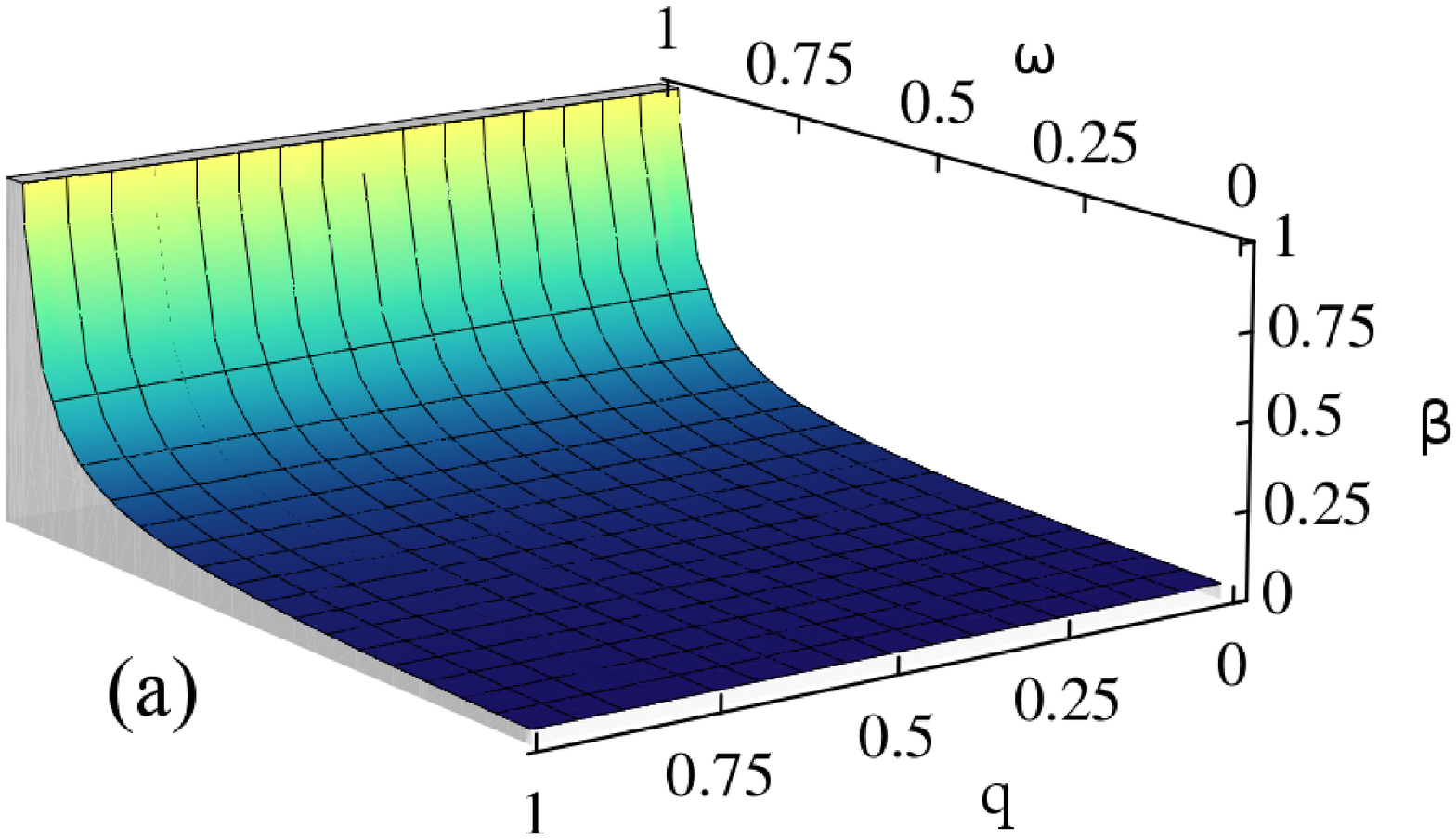}\hspace{0.5cm}
    \includegraphics[width=3.0in]{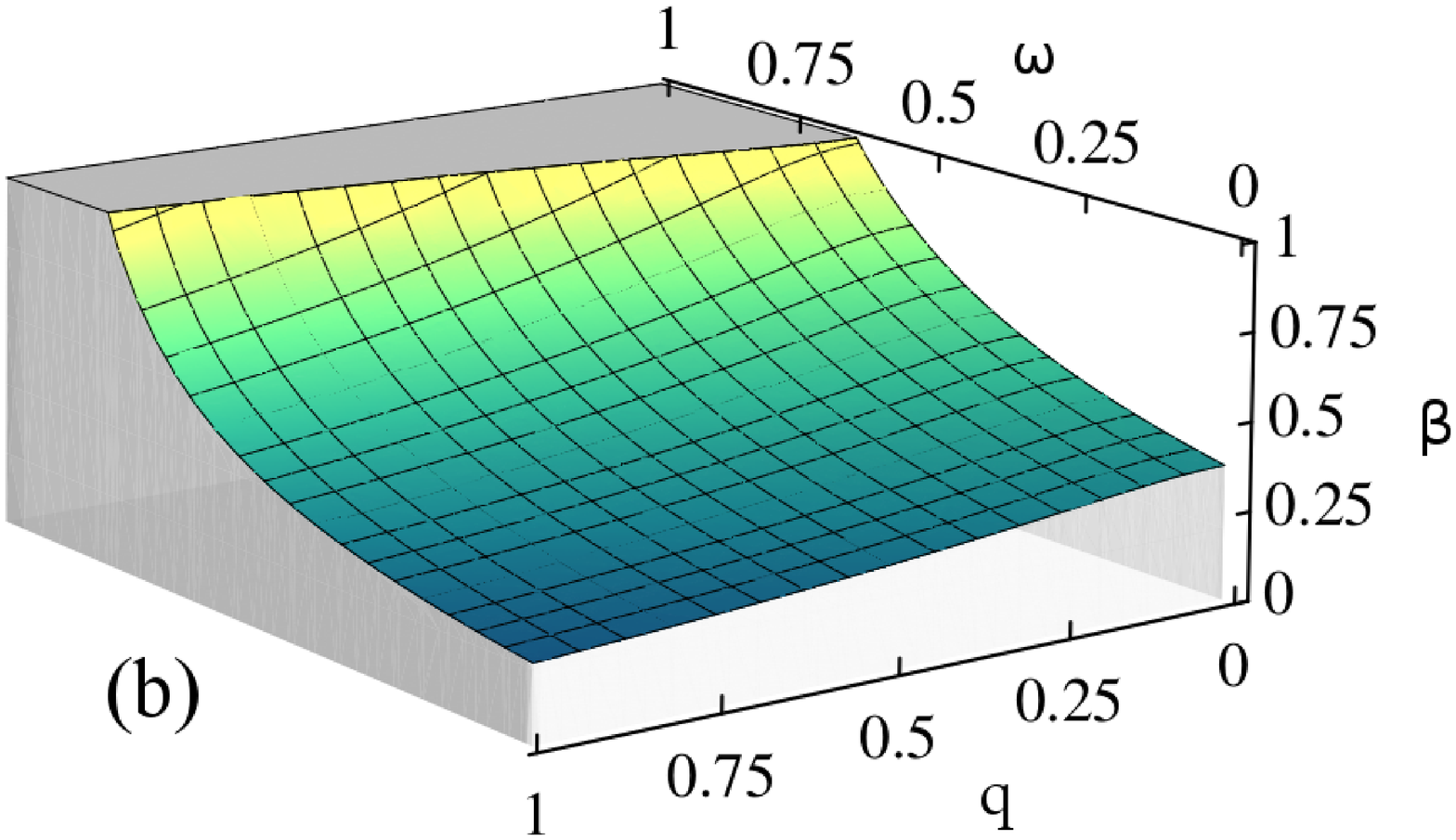}\vspace{1cm}
    \includegraphics[width=3.0in]{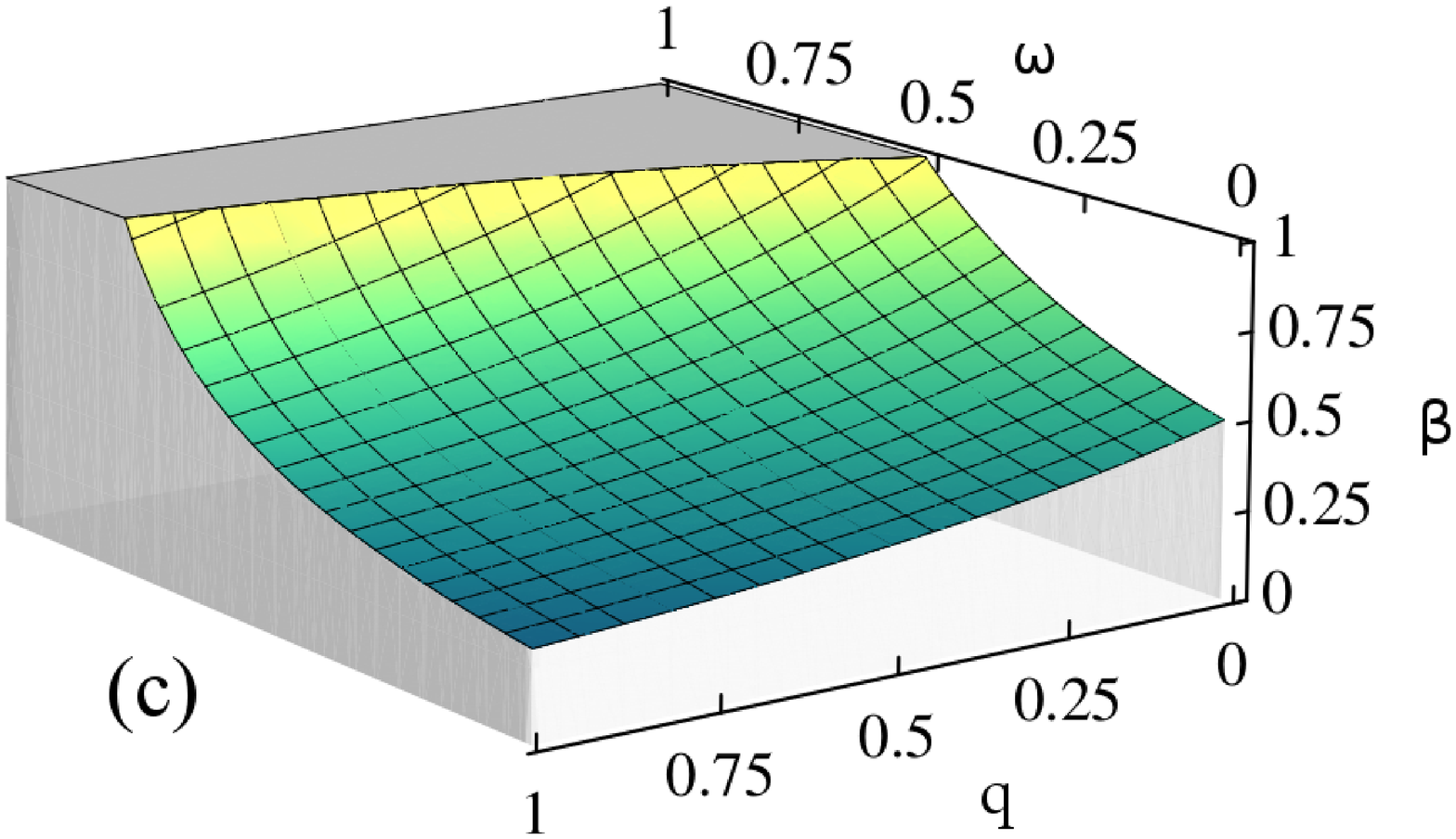}\hspace{0.5cm}
    \includegraphics[width=3.0in]{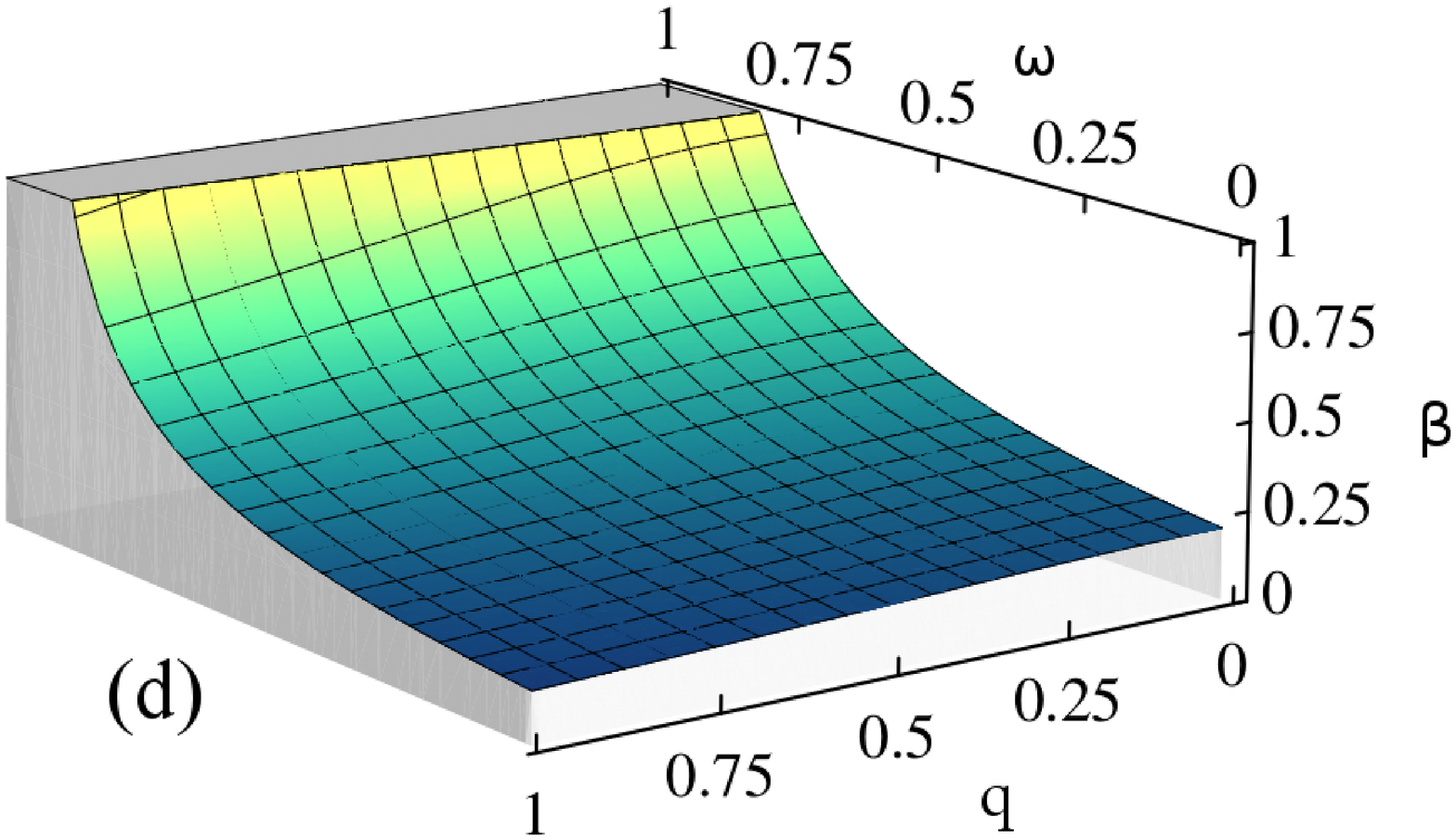}
    \caption{{\bf Surface phase diagram} obtained theoretically for
      the case of $t_r=1$. In the top panels we consider in layer $A$
      an ER network with $\langle k_A \rangle = 2$ with $k_{min}=0$
      and $k_{max}=40$, and in layer $B$ a SF network with (a)
      $\lambda_B = 1.5$ and (b) $\lambda_B = 3.5$, $k_{min}=2$,
      $k_{max}=\sqrt {10^5}$ and exponential cutoff $c = 135$. On the
      bottom we consider two ER networks with $\langle k_A \rangle =
      2$ in layer $A$ and (c) $\langle k_B \rangle =2$ (d) $\langle
      k_B \rangle=5$ in layer $B$. In all $k_{min}=0$ and
      $k_{max}=40$. In all networks, $N=10^5$. We can see that there
      are two well marked regions: an epidemic and a non-epidemic
      phase, separated by a critical value of $\beta$. Increasing the
      overlapping $q$ facilitates the spread of the epidemic, thus,
      $\beta_c$ decreases and the epidemic phase regime is smaller. In
      all figures there is a critical value of $\omega$ above which,
      regardless of the virulence of the disease, there is no
      epidemic. In the case of no vaccination, $\omega=0$, we recover
      the regular SIR in multiplex networks and as $\omega$ increases,
      there will be more vaccinated individuals so $\beta_c$
      increases.}
   \label{diagrama}
  \end{minipage}
\end{figure}

In Fig.~\ref{diagrama} we display a surface plot that shows how the
critical infection probability $\beta_c$ depends on the vaccination
probability $\omega$ and the overlapping between layers $q$. In the figures at the top,
we consider the cases of an ER network in layer $A$, and a SF network
in layer $B$. Figures at the bottom correspond to the case of two ER
networks. In all figures we set $t_r=1$.

From Fig.~(\ref{diagrama}) we can see that the surface, which
represents the critical virulence, separates two phases: one epidemic
(regime above the surface) and another epidemic-free (regime below the
surface). We observe that in all cases as $q$ decreases and $\omega$
increases the non-epidemic phase becomes wider. Furthermore, as we
increase $\omega$, for all values of $q$, we can see the existence of
a threshold above which, even for very virulent diseases, the outbreak
will never become an epidemic. The behavior of the plots in
Fig.~\ref{diagrama} is similar. From the figures above, we can see
that by decreasing $\lambda_B$, we increase the heterogeneity of the
network, and this facilitates the spread of the disease, since once
the infection reaches a hub, a high degree node, it spreads rapidly
across the network. The same applies for figures on the bottom, by
connecting two networks with low average degree, the infection does
not propagate so easily.

\begin{figure}
  \begin{center}
    \includegraphics[width=1\linewidth]{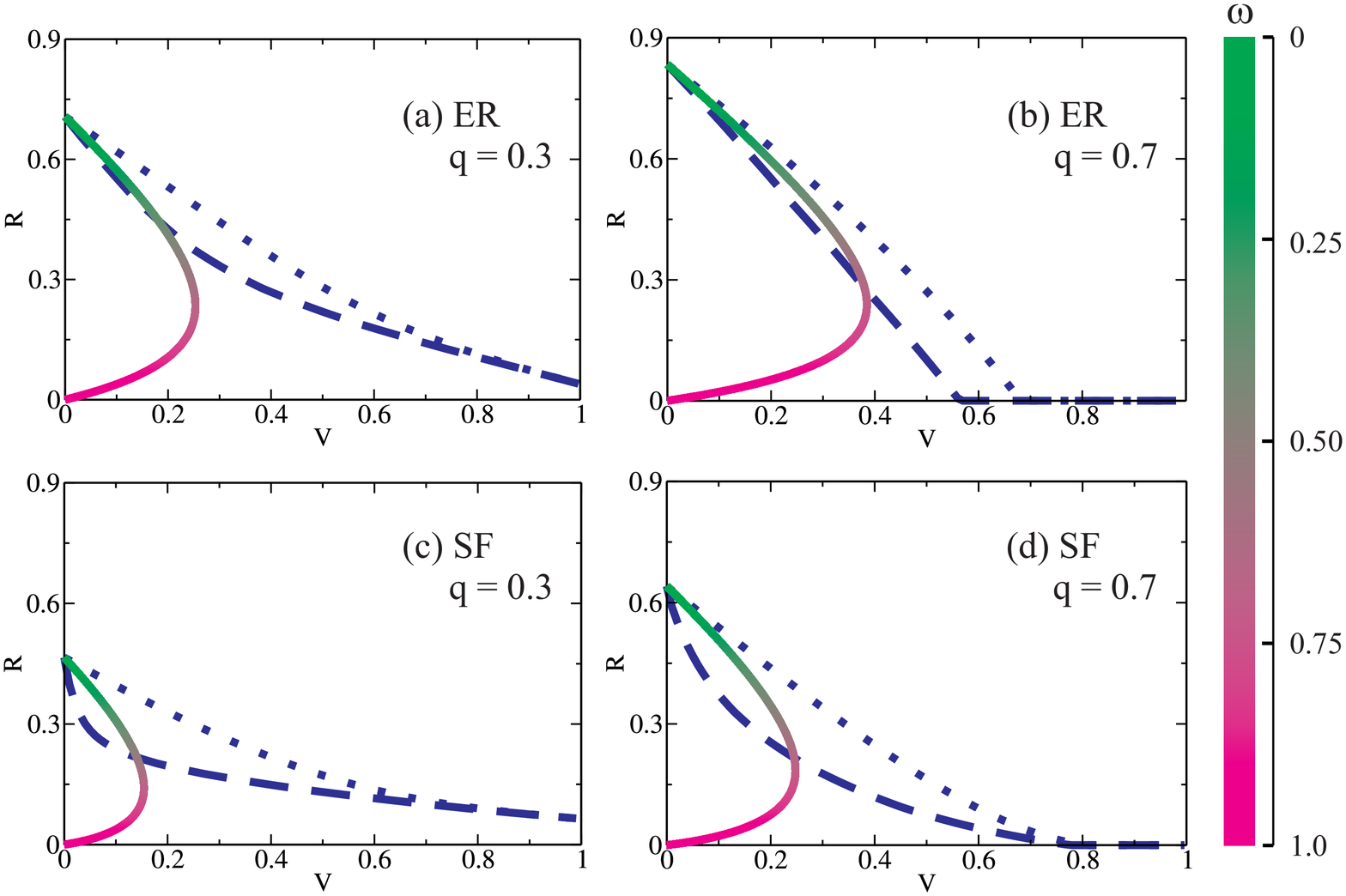}
  \end{center}
  \caption{{\bf Comparison between dynamic, random and targeted
      vaccination}. Theoretical results of the total fraction of
    recovered (R) nodes as a function of V. All dotted lines
    correspond to random vaccination (\protect\blackdotted), the
    dashed lines to targeted immunization (\protect\blackdashdotted)
    and the colored solid lines correspond to our strategy of dynamic
    vaccination. For the case of dynamic vaccination the colors in the
    curves indicate different values of $\omega$. Notice that $\omega$
    decreases from top to bottom.  We set $t_r=1$, $\beta=0.3$ and
    consider different values of overlap (a)-(c) $q=0.3$ and (b)-(d)
    $q=0.7$. In (a)-(b) we used two ER networks with $\langle k_A
    \rangle = \langle k_B \rangle = 5$ with $k_{min}=0$ and
    $k_{max}=40$, and in (c)-(d) two scale free networks, with
    $\lambda_A = \lambda_B = 2.5$, $k_{min}=2$, $k_{max}=\sqrt
            {N}$, $N=10^5$ and exponential cutoff $c = 50$.}
  \label{comparacion}
\end{figure}

Finally, in Fig.~\ref{comparacion} we theoretically compare our
strategy of dynamic vaccination with a very effective strategy,
targeted vaccination \cite{Buono_15}, and with another that usually
performs poorly, random vaccination \cite{alvarez2015}. In the case of
random vaccination, a random fraction $V$ of nodes in layer $A$ are
vaccinated before the spreading of the disease, and if those nodes
have a counterpart in layer $B$, they will also get immunization
against the infection. On the other hand, in the targeted immunization
strategy, we choose a fraction $V$ of the highest connected
individuals in layer $A$ to get vaccinated.

We consider the case of (a)-(b) two ER multiplex networks with the
same average degree, equal to $\langle k_{A/B} \rangle = 5$, and
(c)-(d) two scale free networks where $\lambda_A=\lambda_B=2.5$. We
set $t_r = 1$, $\beta=0.3$, and vary the overlap between networks
(a)-(c) $q =0.3$ and (c)-(d) $q=0.7$. Fig.~\ref{comparacion} shows the
total fraction of recovered nodes in the steady state as a function of
$V$ and $\omega$. In Fig.\ref{comparacion}(a), when the total fraction
of vaccinated individuals is, for instance, $V=0.2$, for the case of
random vaccination $R \simeq 0.55$, while targeted immunization
performs much better, $R \simeq 0.15$. On the other hand, note that
for dynamic vaccination there are two possible outcomes of $R$
depending on $\omega$. Using the same amount of vaccines ($V=0.2$), if
$\omega \simeq 0.3$ then $R \simeq 0.4$, but for $\omega \simeq 0.6$,
we have a much more interesting scenario where the final fraction of
recovered individuals is close $R \simeq 0.1$. In this last case,
there is almost no difference between targeted and dynamic
vaccination. In some cases, there is a threshold value $V^*$, above
which there is no epidemic. For instance, in
Fig.~\ref{comparacion}(b), $V^*=0.45$ (targeted) and $V^*=0.65$
(random). However, in the case of dynamic vaccination $V^* \rightarrow
0$ when $\omega \rightarrow 1$, $i.e.$, a very small amount of
vaccines is needed to prevent the epidemic.
In short, from Fig.~\ref{comparacion} we can see that, for the same
amount of vaccinated individuals, the number of recovered individuals
is significantly lower for dynamic vaccination compared to random
vaccination. This is observed in all plots, which indicates that our
strategy is significantly more effective than random
vaccination. Especially, in more heterogeneous networks, such as SF
networks, we can see that this strategy is even more efficient.

On the other hand, if we compare our strategy with targeted
vaccination, there is a certain value of $\omega$ above which in
dynamic vaccination performs much better, even using a lower amount of
vaccines than in targeted vaccination. Notably, this is also observed
for scale free networks, in which the immunization of the hubs is
usually an extremely effective strategy. This is because when the
contacts of an infected node are immunized, is highly likely that at
least one of that contacts is a hub. Thus, for SF networks our
strategy also targets high connectivity nodes, preventing a
massive spreading of the disease.

Furthermore, when the system is more interconnected, the disease
spreads more. Therefore, more vaccines are needed to prevented it from
becoming an epidemic. For example, for a final fraction of recovered
individuals equal to $20\%$ of the population, it is required to
vaccinate a fraction of individuals equal to $V(q=0.3) \simeq 0.25$
and $V(q=0.7) \simeq 0.4$.

To summarize, we demonstrate that, on top of a multiplex network
structure, dynamic vaccination strategy is much more efficient than
random immunization. Using the same amount of vaccines, the total
fraction of recovered individuals is always lower in the case of
dynamic vaccination. Besides, in our strategy, depending on the
parameters, we obtain a region where regardless of the virulence of
the disease it will never become an epidemic. In comparison, if we
vaccinate the same number of individuals, the fraction recovered in
dynamic immunization will always be below that obtained if we
vaccinate at random. Due to this result, our strategy can be very
beneficial to be implemented in real scenarios, for example in
outbreaks of Ebola or influenza, such as H1N1.

\section{Discussion}
\vspace{-0.3cm}

In summary, we studied a novel model of dynamic vaccination on a
system composed of two partially overlapped networks, where $q$ is the
fraction of common nodes in both networks. In our model, susceptible
individuals in contact with infected patients have the opportunity to
be vaccinated before the their neighbors attempt to infect them. That
is, each time an infected person comes in contact with a susceptible
individual, this one will try to get vaccinated with probability
$\omega$ and if he does not succeed he will be infected with
probability $\beta$. Each infected node is assumed to recover after
$t_r$ time steps and will become immunized. Besides, vaccinated nodes
are also immunized and can not be infected or infect others. Mapping
this process into bond percolation and using the framework of
generating functions, we analyzed analytically the total fraction of
recovered and vaccinated nodes in the steady state as a function of
the virulence of the diseases $\beta$ for different values of the
$\omega$ and $q$, and we found a perfect agreement between the
theoretical and the simulation results. As expected we find that as
$\omega$ increases the epidemic threshold $\beta_c$ becomes larger,
and disappearing for very large values of $\omega$. We also find a
peak, for certain values of the parameters, in the fraction of
vaccinated nodes as a function of $\omega$, which is determined by the
competition between the vaccination strategy and the spread of the
disease. We find an interesting phase diagram in the plane $\beta -
\omega$, where we can see an epidemic phase, which diminish as $q$ and
$\omega$ increase, and a non-epidemic phase, where the diseases can
not spread. A remarkable result of the phase diagram is that for
certain values of $q$ and $\omega$, regardless of the virulence of the
disease, it will never become an epidemic. Finally, we demonstrate
that our strategy is always more efficient than random immunization
and depending on $\omega$, performs better or worse than targeted
vaccination.  For high vaccination probabilities, i.e., when the
population is more receptive to be vaccinated, dynamic immunization is
the most effective strategy to avoid or mitigate an epidemic. This
goal can be achieved, using efficiently the available vaccines,
immunizing a small fraction of the population and creating a barrier
of vaccinated individuals that the infection can not pass through.

\section{Acknowledgments}

SH thanks the Israel Science Foundation, ONR, the Israel Ministry of
Science and Technology (MOST) with the Italy Ministry of Foreign
Affairs, BSF-NSF, MOST with the Japan Science and Technology Agency,
the BIU Center for Research in Applied Cryptography and Cyber
Security, and DTRA (Grant no. HDTRA-1-10-1- 0014) for financial
support. LGAZ, MAD and LAB wish to thank to UNMdP, FONCyT and CONICET
(Pict 0429/2013, Pict 1407/2014 and PIP 00443/2014) for financial
support.


\section*{Appendix}

\subsection{Dependence on $T_{\beta}$}
\label{appen1}

In this section we show the dependence on the overall transmissibility
$T_{\beta}$ of the recovered and vaccinated fraction of nodes, $R$ and
$V$, in the steady state. Also, we show the dependence on the
prefactors that multiply Eqs. (\ref{R}) and (\ref{V}), which measure
the relative weight of the probabilities of infection (or vaccination)
in the dynamic.

As we saw earlier in Sec. \ref{theory}, from
Eqs. (\ref{transmissibility}) and (\ref{transmissibility2}) we find
the prefactors in Eqs. (\ref{R}) and (\ref{V}) as,
\begin{eqnarray}
  B_{\beta} &=& \frac{T_{\beta}}{T_{\beta}+T_{\omega}}  = \frac{(1-\omega)\;\beta}{(1-\omega)\beta+\omega}, \nonumber\\
  B_{\omega} &=& \frac{T_{\omega}}{T_{\beta}+T_{\omega}} = \frac{\omega}{(1-\omega)\beta+\omega}.
  \label{factor}
\end{eqnarray}
Note, interestingly these prefactors are independent of the recovery
time $t_r$.

\begin{figure}[!t]
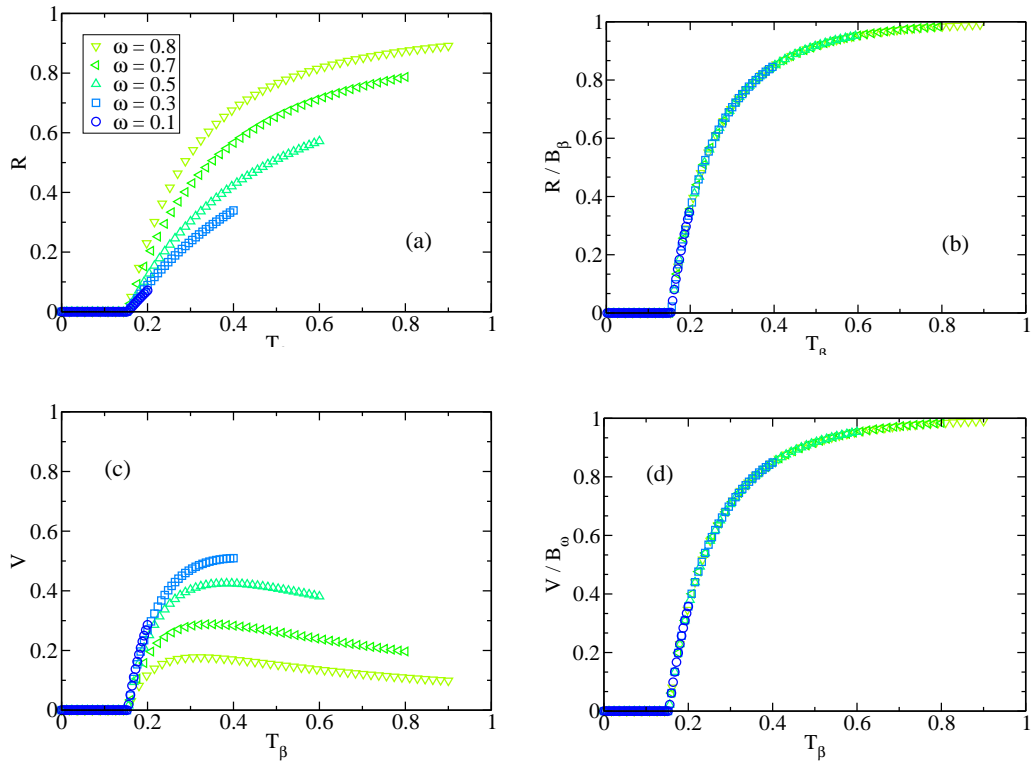

  \begin{center}
    \includegraphics[width=0.4\linewidth]{6a.eps}\hspace{0.5cm}
    \includegraphics[width=0.4\linewidth]{6b.eps}\vspace{0.5cm}
    \includegraphics[width=0.4\linewidth]{6c.eps}\hspace{0.5cm}
    \includegraphics[width=0.4\linewidth]{6d.eps}
  \end{center}
  \caption{{\bf Theoretical Results} of the total fraction of (a)
    recovered $R$, (b) $R/B_{\beta}$, (c) vaccinated $V$ and (d)
    $V/B_{\omega}$ nodes as a function of the overall transmissibility
    $T_{\beta}=(1-\omega)\;\beta$ when $t_r=1$ and for different
    values of $\omega$, in the steady state. The multiplex network
    consists of two ER networks and we set $q=0.3$. Each network size
    is $N= 10^5$ and $\langle k_A \rangle = \langle k_B \rangle = 5$
    with $k_{min}=0$ and $k_{max}= 40$.}
  \label{Rec-solp}
\end{figure}

In Fig.~\ref{Rec-solp} we plot in (a) the total fraction of recovered
nodes $R$, (b) $R/B_{\beta}$, (c) total fraction of vaccinated nodes
$V$ and (d) $V/B_{\omega}$, as a function of the epidemic
transmissibility $T_{\beta}$ for different values of $\omega$, from
$\omega=0.1$ to $\omega=0.8$, with intervals of $\Delta \omega =
0.1$. For the sake of simplicity we consider the case $t_r=1$. Thus,
from Eq. (\ref{transmissibility}) the epidemic transmissibility is
reduced to $T_{\beta} = (1-\omega) \beta$.

The multiplex network consists of two coupled Erd\H{o}s-R\'enyi (ER)
networks with $q=0.3$, average degree $\langle k_A \rangle = \langle
k_B \rangle = 5$ and, $k_{min}=0$ and $k_{max}=40$.

As we can see, in Fig.~\ref{Rec-solp}(a) and (c) for the different
$\omega$ values, when $R$ and $V$ are plotted as a function of
$T_{\beta}$ all the curves collapse in the same critical threshold,
when $\omega=0$ \cite{Buono_15}, which is equal to,
\begin{equation}
  T_{\beta}|_{(\beta=\beta_c)} = \frac{1}{\langle k \rangle \: (1+q)}.
\end{equation}

Since we consider $\langle k \rangle=\langle k_A \rangle=\langle k_B
\rangle=5$ and $q=0.3$, then for Fig.~\ref{Rec-solp} the critical
threshold is $T_{\beta}|_{(\beta=\beta_c)} \simeq 0.15$.

Notice that, for instance, for $t_r=1$ we have $T_\beta=\beta \;
(1-\omega)$, and thus for a fixed vaccination probability $T_\beta \in
[0,1-\omega]$, with $\omega \in [0,1]$. Then, $T_\beta \notin [0,1]$,
as the transmissibility in the standard SIR.  In
Figure~\ref{Rec-solp}, as each curve corresponds to a different value
of $\omega$, they will reach maximum values at different
$T_\beta$. Furthermore, as $\omega$ increases each curve becomes
shorter since the dependency on $\omega$ is still present in the
prefactors.

In Fig.~\ref{Rec-solp} (b) and (d) we multiply all the curves
(fraction of recovered nodes) by the weighted scaled prefactor of
infection, and we can see that all data collapse into a single
curve. This is due to the fact that with this scaling the dependence
on $\omega$ is removed.


\begin{thebibliography}{63}%
\makeatletter
\providecommand \@ifxundefined [1]{%
 \@ifx{#1\undefined}
}%
\providecommand \@ifnum [1]{%
 \ifnum #1\expandafter \@firstoftwo
 \else \expandafter \@secondoftwo
 \fi
}%
\providecommand \@ifx [1]{%
 \ifx #1\expandafter \@firstoftwo
 \else \expandafter \@secondoftwo
 \fi
}%
\providecommand \natexlab [1]{#1}%
\providecommand \enquote  [1]{``#1''}%
\providecommand \bibnamefont  [1]{#1}%
\providecommand \bibfnamefont [1]{#1}%
\providecommand \citenamefont [1]{#1}%
\providecommand \href@noop [0]{\@secondoftwo}%
\providecommand \href [0]{\begingroup \@sanitize@url \@href}%
\providecommand \@href[1]{\@@startlink{#1}\@@href}%
\providecommand \@@href[1]{\endgroup#1\@@endlink}%
\providecommand \@sanitize@url [0]{\catcode `\\12\catcode `\$12\catcode
  `\&12\catcode `\#12\catcode `\^12\catcode `\_12\catcode `\%12\relax}%
\providecommand \@@startlink[1]{}%
\providecommand \@@endlink[0]{}%
\providecommand \url  [0]{\begingroup\@sanitize@url \@url }%
\providecommand \@url [1]{\endgroup\@href {#1}{\urlprefix }}%
\providecommand \urlprefix  [0]{URL }%
\providecommand \Eprint [0]{\href }%
\providecommand \doibase [0]{http://dx.doi.org/}%
\providecommand \selectlanguage [0]{\@gobble}%
\providecommand \bibinfo  [0]{\@secondoftwo}%
\providecommand \bibfield  [0]{\@secondoftwo}%
\providecommand \translation [1]{[#1]}%
\providecommand \BibitemOpen [0]{}%
\providecommand \bibitemStop [0]{}%
\providecommand \bibitemNoStop [0]{.\EOS\space}%
\providecommand \EOS [0]{\spacefactor3000\relax}%
\providecommand \BibitemShut  [1]{\csname bibitem#1\endcsname}%
\let\auto@bib@innerbib\@empty
\bibitem [{\citenamefont {Fraser}\ \emph {et~al.}(2009)\citenamefont {Fraser},
  \citenamefont {Donnelly}, \citenamefont {Cauchemez}, \citenamefont {Hanage},
  \citenamefont {Van~Kerkhove}, \citenamefont {Hollingsworth}, \citenamefont
  {Griffin}, \citenamefont {Baggaley}, \citenamefont {Jenkins}, \citenamefont
  {Lyons} \emph {et~al.}}]{H1N1_fraser}%
  \BibitemOpen
  \bibfield  {author} {\bibinfo {author} {\bibfnamefont {C.}~\bibnamefont
  {Fraser}}, \bibinfo {author} {\bibfnamefont {C.~A.}\ \bibnamefont
  {Donnelly}}, \bibinfo {author} {\bibfnamefont {S.}~\bibnamefont {Cauchemez}},
  \bibinfo {author} {\bibfnamefont {W.~P.}\ \bibnamefont {Hanage}}, \bibinfo
  {author} {\bibfnamefont {M.~D.}\ \bibnamefont {Van~Kerkhove}}, \bibinfo
  {author} {\bibfnamefont {T.~D.}\ \bibnamefont {Hollingsworth}}, \bibinfo
  {author} {\bibfnamefont {J.}~\bibnamefont {Griffin}}, \bibinfo {author}
  {\bibfnamefont {R.~F.}\ \bibnamefont {Baggaley}}, \bibinfo {author}
  {\bibfnamefont {H.~E.}\ \bibnamefont {Jenkins}}, \bibinfo {author}
  {\bibfnamefont {E.~J.}\ \bibnamefont {Lyons}},  \emph {et~al.},\ }\href@noop
  {} {\bibfield  {journal} {\bibinfo  {journal} {science}\ }\textbf {\bibinfo
  {volume} {324}},\ \bibinfo {pages} {1557} (\bibinfo {year}
  {2009})}\BibitemShut {NoStop}%
\bibitem [{\citenamefont {Pourbohloul}\ \emph {et~al.}(2009)\citenamefont
  {Pourbohloul}, \citenamefont {Ahued}, \citenamefont {Davoudi}, \citenamefont
  {Meza}, \citenamefont {Meyers}, \citenamefont {Skowronski}, \citenamefont
  {Villase{\~n}or}, \citenamefont {Galv{\'a}n}, \citenamefont {Cravioto},
  \citenamefont {Earn} \emph {et~al.}}]{H1N1_meyers}%
  \BibitemOpen
  \bibfield  {author} {\bibinfo {author} {\bibfnamefont {B.}~\bibnamefont
  {Pourbohloul}}, \bibinfo {author} {\bibfnamefont {A.}~\bibnamefont {Ahued}},
  \bibinfo {author} {\bibfnamefont {B.}~\bibnamefont {Davoudi}}, \bibinfo
  {author} {\bibfnamefont {R.}~\bibnamefont {Meza}}, \bibinfo {author}
  {\bibfnamefont {L.~A.}\ \bibnamefont {Meyers}}, \bibinfo {author}
  {\bibfnamefont {D.~M.}\ \bibnamefont {Skowronski}}, \bibinfo {author}
  {\bibfnamefont {I.}~\bibnamefont {Villase{\~n}or}}, \bibinfo {author}
  {\bibfnamefont {F.}~\bibnamefont {Galv{\'a}n}}, \bibinfo {author}
  {\bibfnamefont {P.}~\bibnamefont {Cravioto}}, \bibinfo {author}
  {\bibfnamefont {D.~J.}\ \bibnamefont {Earn}},  \emph {et~al.},\ }\href@noop
  {} {\bibfield  {journal} {\bibinfo  {journal} {Influenza and other
  respiratory viruses}\ }\textbf {\bibinfo {volume} {3}},\ \bibinfo {pages}
  {215} (\bibinfo {year} {2009})}\BibitemShut {NoStop}%
\bibitem [{\citenamefont {White}\ \emph {et~al.}(2009)\citenamefont {White},
  \citenamefont {Wallinga}, \citenamefont {Finelli}, \citenamefont {Reed},
  \citenamefont {Riley}, \citenamefont {Lipsitch},\ and\ \citenamefont
  {Pagano}}]{H1N1_r0}%
  \BibitemOpen
  \bibfield  {author} {\bibinfo {author} {\bibfnamefont {L.~F.}\ \bibnamefont
  {White}}, \bibinfo {author} {\bibfnamefont {J.}~\bibnamefont {Wallinga}},
  \bibinfo {author} {\bibfnamefont {L.}~\bibnamefont {Finelli}}, \bibinfo
  {author} {\bibfnamefont {C.}~\bibnamefont {Reed}}, \bibinfo {author}
  {\bibfnamefont {S.}~\bibnamefont {Riley}}, \bibinfo {author} {\bibfnamefont
  {M.}~\bibnamefont {Lipsitch}}, \ and\ \bibinfo {author} {\bibfnamefont
  {M.}~\bibnamefont {Pagano}},\ }\href@noop {} {\bibfield  {journal} {\bibinfo
  {journal} {Influenza and other respiratory viruses}\ }\textbf {\bibinfo
  {volume} {3}},\ \bibinfo {pages} {267} (\bibinfo {year} {2009})}\BibitemShut
  {NoStop}%
\bibitem [{\citenamefont {Riley}\ \emph {et~al.}(2003)\citenamefont {Riley},
  \citenamefont {Fraser}, \citenamefont {Donnelly}, \citenamefont {Ghani},
  \citenamefont {Abu-Raddad}, \citenamefont {Hedley}, \citenamefont {Leung},
  \citenamefont {Ho}, \citenamefont {Lam}, \citenamefont {Thach} \emph
  {et~al.}}]{riley_03}%
  \BibitemOpen
  \bibfield  {author} {\bibinfo {author} {\bibfnamefont {S.}~\bibnamefont
  {Riley}}, \bibinfo {author} {\bibfnamefont {C.}~\bibnamefont {Fraser}},
  \bibinfo {author} {\bibfnamefont {C.~A.}\ \bibnamefont {Donnelly}}, \bibinfo
  {author} {\bibfnamefont {A.~C.}\ \bibnamefont {Ghani}}, \bibinfo {author}
  {\bibfnamefont {L.~J.}\ \bibnamefont {Abu-Raddad}}, \bibinfo {author}
  {\bibfnamefont {A.~J.}\ \bibnamefont {Hedley}}, \bibinfo {author}
  {\bibfnamefont {G.~M.}\ \bibnamefont {Leung}}, \bibinfo {author}
  {\bibfnamefont {L.-M.}\ \bibnamefont {Ho}}, \bibinfo {author} {\bibfnamefont
  {T.-H.}\ \bibnamefont {Lam}}, \bibinfo {author} {\bibfnamefont {T.~Q.}\
  \bibnamefont {Thach}},  \emph {et~al.},\ }\href@noop {} {\bibfield  {journal}
  {\bibinfo  {journal} {Science}\ }\textbf {\bibinfo {volume} {300}},\ \bibinfo
  {pages} {1961} (\bibinfo {year} {2003})}\BibitemShut {NoStop}%
\bibitem [{\citenamefont {Meyers}\ \emph {et~al.}(2005)\citenamefont {Meyers},
  \citenamefont {Pourbohloul}, \citenamefont {Newman}, \citenamefont
  {Skowronski},\ and\ \citenamefont {Brunham}}]{sars_meyers}%
  \BibitemOpen
  \bibfield  {author} {\bibinfo {author} {\bibfnamefont {L.~A.}\ \bibnamefont
  {Meyers}}, \bibinfo {author} {\bibfnamefont {B.}~\bibnamefont {Pourbohloul}},
  \bibinfo {author} {\bibfnamefont {M.~E.}\ \bibnamefont {Newman}}, \bibinfo
  {author} {\bibfnamefont {D.~M.}\ \bibnamefont {Skowronski}}, \ and\ \bibinfo
  {author} {\bibfnamefont {R.~C.}\ \bibnamefont {Brunham}},\ }\href@noop {}
  {\bibfield  {journal} {\bibinfo  {journal} {Journal of theoretical biology}\
  }\textbf {\bibinfo {volume} {232}},\ \bibinfo {pages} {71} (\bibinfo {year}
  {2005})}\BibitemShut {NoStop}%
\bibitem [{\citenamefont {Colizza}\ \emph {et~al.}(2007)\citenamefont
  {Colizza}, \citenamefont {Barrat}, \citenamefont {Barth{\'e}lemy},\ and\
  \citenamefont {Vespignani}}]{Col_07}%
  \BibitemOpen
  \bibfield  {author} {\bibinfo {author} {\bibfnamefont {V.}~\bibnamefont
  {Colizza}}, \bibinfo {author} {\bibfnamefont {A.}~\bibnamefont {Barrat}},
  \bibinfo {author} {\bibfnamefont {M.}~\bibnamefont {Barth{\'e}lemy}}, \ and\
  \bibinfo {author} {\bibfnamefont {A.}~\bibnamefont {Vespignani}},\
  }\href@noop {} {\bibfield  {journal} {\bibinfo  {journal} {BMC Medicine}\
  }\textbf {\bibinfo {volume} {5}},\ \bibinfo {pages} {34} (\bibinfo {year}
  {2007})}\BibitemShut {NoStop}%
\bibitem [{\citenamefont {Bajardi}\ \emph {et~al.}(2011)\citenamefont
  {Bajardi}, \citenamefont {Poletto}, \citenamefont {Ramasco}, \citenamefont
  {Tizzoni}, \citenamefont {Colizza},\ and\ \citenamefont
  {Vespignani}}]{Colizza_11}%
  \BibitemOpen
  \bibfield  {author} {\bibinfo {author} {\bibfnamefont {P.}~\bibnamefont
  {Bajardi}}, \bibinfo {author} {\bibfnamefont {C.}~\bibnamefont {Poletto}},
  \bibinfo {author} {\bibfnamefont {J.~J.}\ \bibnamefont {Ramasco}}, \bibinfo
  {author} {\bibfnamefont {M.}~\bibnamefont {Tizzoni}}, \bibinfo {author}
  {\bibfnamefont {V.}~\bibnamefont {Colizza}}, \ and\ \bibinfo {author}
  {\bibfnamefont {A.}~\bibnamefont {Vespignani}},\ }\href {\doibase
  10.1371/journal.pone.0016591} {\bibfield  {journal} {\bibinfo  {journal}
  {PLOS ONE}\ }\textbf {\bibinfo {volume} {6}},\ \bibinfo {pages} {1} (\bibinfo
  {year} {2011})}\BibitemShut {NoStop}%
\bibitem [{\citenamefont {Anderson}\ and\ \citenamefont
  {May}(1992)}]{Ander_91}%
  \BibitemOpen
  \bibfield  {author} {\bibinfo {author} {\bibfnamefont {R.~M.}\ \bibnamefont
  {Anderson}}\ and\ \bibinfo {author} {\bibfnamefont {R.~M.}\ \bibnamefont
  {May}},\ }\href@noop {} {\emph {\bibinfo {title} {{Infectious Diseases of
  Humans: Dynamics and Control}}}}\ (\bibinfo  {publisher} {Oxford University
  Press, Oxford},\ \bibinfo {year} {1992})\BibitemShut {NoStop}%
\bibitem [{\citenamefont {Bailey}(1975)}]{Bailey_75}%
  \BibitemOpen
  \bibfield  {author} {\bibinfo {author} {\bibfnamefont {N.~T.~J.}\
  \bibnamefont {Bailey}},\ }\href@noop {} {\emph {\bibinfo {title} {{The
  Mathematical Theory of Infectious Diseases}}}}\ (\bibinfo  {publisher}
  {Griffin, London},\ \bibinfo {year} {1975})\BibitemShut {NoStop}%
\bibitem [{\citenamefont {Cattuto}\ \emph {et~al.}(2010)\citenamefont
  {Cattuto}, \citenamefont {den Broeck}, \citenamefont {Barrat}, \citenamefont
  {Colizza}, \citenamefont {Pinton},\ and\ \citenamefont
  {Vespignani}}]{Catt_01}%
  \BibitemOpen
  \bibfield  {author} {\bibinfo {author} {\bibfnamefont {C.}~\bibnamefont
  {Cattuto}}, \bibinfo {author} {\bibfnamefont {W.~V.}\ \bibnamefont {den
  Broeck}}, \bibinfo {author} {\bibfnamefont {A.}~\bibnamefont {Barrat}},
  \bibinfo {author} {\bibfnamefont {V.}~\bibnamefont {Colizza}}, \bibinfo
  {author} {\bibfnamefont {J.-F.}\ \bibnamefont {Pinton}}, \ and\ \bibinfo
  {author} {\bibfnamefont {A.}~\bibnamefont {Vespignani}},\ }\href@noop {}
  {\bibfield  {journal} {\bibinfo  {journal} {PLoS ONE}\ }\textbf {\bibinfo
  {volume} {5}},\ \bibinfo {pages} {e11596} (\bibinfo {year}
  {2010})}\BibitemShut {NoStop}%
\bibitem [{\citenamefont {Gonzalez}\ \emph {et~al.}(2008)\citenamefont
  {Gonzalez}, \citenamefont {Hidalgo},\ and\ \citenamefont
  {Barabasi}}]{gonzalez_08}%
  \BibitemOpen
  \bibfield  {author} {\bibinfo {author} {\bibfnamefont {M.~C.}\ \bibnamefont
  {Gonzalez}}, \bibinfo {author} {\bibfnamefont {C.~A.}\ \bibnamefont
  {Hidalgo}}, \ and\ \bibinfo {author} {\bibfnamefont {A.-L.}\ \bibnamefont
  {Barabasi}},\ }\href@noop {} {\bibfield  {journal} {\bibinfo  {journal}
  {Nature}\ }\textbf {\bibinfo {volume} {453}},\ \bibinfo {pages} {779}
  (\bibinfo {year} {2008})}\BibitemShut {NoStop}%
\bibitem [{\citenamefont {G{\'o}mez-Garde{\~n}es}\ \emph
  {et~al.}(2008)\citenamefont {G{\'o}mez-Garde{\~n}es}, \citenamefont {Latora},
  \citenamefont {Moreno},\ and\ \citenamefont {Profumo}}]{gardenes_08}%
  \BibitemOpen
  \bibfield  {author} {\bibinfo {author} {\bibfnamefont {J.}~\bibnamefont
  {G{\'o}mez-Garde{\~n}es}}, \bibinfo {author} {\bibfnamefont {V.}~\bibnamefont
  {Latora}}, \bibinfo {author} {\bibfnamefont {Y.}~\bibnamefont {Moreno}}, \
  and\ \bibinfo {author} {\bibfnamefont {E.}~\bibnamefont {Profumo}},\
  }\href@noop {} {\bibfield  {journal} {\bibinfo  {journal} {Proceedings of the
  National Academy of Sciences}\ }\textbf {\bibinfo {volume} {105}},\ \bibinfo
  {pages} {1399} (\bibinfo {year} {2008})}\BibitemShut {NoStop}%
\bibitem [{\citenamefont {Boccaletti}\ \emph {et~al.}(2006)\citenamefont
  {Boccaletti}, \citenamefont {Latora}, \citenamefont {Moreno}, \citenamefont
  {Chavez},\ and\ \citenamefont {Hwang}}]{Boc_01}%
  \BibitemOpen
  \bibfield  {author} {\bibinfo {author} {\bibfnamefont {S.}~\bibnamefont
  {Boccaletti}}, \bibinfo {author} {\bibfnamefont {V.}~\bibnamefont {Latora}},
  \bibinfo {author} {\bibfnamefont {Y.}~\bibnamefont {Moreno}}, \bibinfo
  {author} {\bibfnamefont {M.}~\bibnamefont {Chavez}}, \ and\ \bibinfo {author}
  {\bibfnamefont {D.}~\bibnamefont {Hwang}},\ }\href@noop {} {\bibfield
  {journal} {\bibinfo  {journal} {Physics Reports}\ }\textbf {\bibinfo {volume}
  {424}},\ \bibinfo {pages} {175} (\bibinfo {year} {2006})}\BibitemShut
  {NoStop}%
\bibitem [{\citenamefont {Cohen}\ and\ \citenamefont {Havlin}(2010)}]{Coh_10}%
  \BibitemOpen
  \bibfield  {author} {\bibinfo {author} {\bibfnamefont {R.}~\bibnamefont
  {Cohen}}\ and\ \bibinfo {author} {\bibfnamefont {S.}~\bibnamefont {Havlin}},\
  }\href@noop {} {\emph {\bibinfo {title} {Complex Networks: Structure,
  Robustness and Function}}}\ (\bibinfo  {publisher} {Cambridge University
  Press},\ \bibinfo {year} {2010})\BibitemShut {NoStop}%
\bibitem [{\citenamefont {Barrat}\ \emph {et~al.}(2004)\citenamefont {Barrat},
  \citenamefont {Barth{\'e}lemy}, \citenamefont {Pastor-Satorras},\ and\
  \citenamefont {Vespignani}}]{Barrat_04}%
  \BibitemOpen
  \bibfield  {author} {\bibinfo {author} {\bibfnamefont {A.}~\bibnamefont
  {Barrat}}, \bibinfo {author} {\bibfnamefont {M.}~\bibnamefont
  {Barth{\'e}lemy}}, \bibinfo {author} {\bibfnamefont {R.}~\bibnamefont
  {Pastor-Satorras}}, \ and\ \bibinfo {author} {\bibfnamefont {A.}~\bibnamefont
  {Vespignani}},\ }\href@noop {} {\bibfield  {journal} {\bibinfo  {journal}
  {Proc. Natl. Acad. Sci. USA}\ }\textbf {\bibinfo {volume} {101}},\ \bibinfo
  {pages} {3747} (\bibinfo {year} {2004})}\BibitemShut {NoStop}%
\bibitem [{\citenamefont {Newman}(2010)}]{New_10}%
  \BibitemOpen
  \bibfield  {author} {\bibinfo {author} {\bibfnamefont {M.~E.~J.}\
  \bibnamefont {Newman}},\ }\href@noop {} {\emph {\bibinfo {title} {Networks:
  An Introduction}}}\ (\bibinfo  {publisher} {Oxford University Press},\
  \bibinfo {year} {2010})\BibitemShut {NoStop}%
\bibitem [{\citenamefont {Pastor-Satorras}\ and\ \citenamefont
  {Vespignani}(2001)}]{past_01}%
  \BibitemOpen
  \bibfield  {author} {\bibinfo {author} {\bibfnamefont {R.}~\bibnamefont
  {Pastor-Satorras}}\ and\ \bibinfo {author} {\bibfnamefont {A.}~\bibnamefont
  {Vespignani}},\ }\href@noop {} {\bibfield  {journal} {\bibinfo  {journal}
  {Phys Rev Lett}\ }\textbf {\bibinfo {volume} {86}},\ \bibinfo {pages} {3200}
  (\bibinfo {year} {2001})}\BibitemShut {NoStop}%
\bibitem [{\citenamefont {Newman}(2002)}]{New_05}%
  \BibitemOpen
  \bibfield  {author} {\bibinfo {author} {\bibfnamefont {M.~E.~J.}\
  \bibnamefont {Newman}},\ }\href@noop {} {\bibfield  {journal} {\bibinfo
  {journal} {Physical Review E}\ }\textbf {\bibinfo {volume} {66}},\ \bibinfo
  {pages} {016128} (\bibinfo {year} {2002})}\BibitemShut {NoStop}%
\bibitem [{\citenamefont {Moreno}\ \emph {et~al.}(2002)\citenamefont {Moreno},
  \citenamefont {Pastor-Satorras},\ and\ \citenamefont
  {Vespignani}}]{moreno_02}%
  \BibitemOpen
  \bibfield  {author} {\bibinfo {author} {\bibfnamefont {Y.}~\bibnamefont
  {Moreno}}, \bibinfo {author} {\bibfnamefont {R.}~\bibnamefont
  {Pastor-Satorras}}, \ and\ \bibinfo {author} {\bibfnamefont {A.}~\bibnamefont
  {Vespignani}},\ }\href@noop {} {\bibfield  {journal} {\bibinfo  {journal}
  {The European Physical Journal B-Condensed Matter and Complex Systems}\
  }\textbf {\bibinfo {volume} {26}},\ \bibinfo {pages} {521} (\bibinfo {year}
  {2002})}\BibitemShut {NoStop}%
\bibitem [{\citenamefont {Valdez}\ \emph {et~al.}(2012)\citenamefont {Valdez},
  \citenamefont {Macri},\ and\ \citenamefont {Braunstein}}]{Val_12}%
  \BibitemOpen
  \bibfield  {author} {\bibinfo {author} {\bibfnamefont {L.~D.}\ \bibnamefont
  {Valdez}}, \bibinfo {author} {\bibfnamefont {P.~A.}\ \bibnamefont {Macri}}, \
  and\ \bibinfo {author} {\bibfnamefont {L.~A.}\ \bibnamefont {Braunstein}},\
  }\href {\doibase 10.1371/journal.pone.0044188} {\bibfield  {journal}
  {\bibinfo  {journal} {PLOS ONE}\ }\textbf {\bibinfo {volume} {7}},\ \bibinfo
  {pages} {1} (\bibinfo {year} {2012})}\BibitemShut {NoStop}%
\bibitem [{\citenamefont {Callaway}\ \emph {et~al.}(2000)\citenamefont
  {Callaway}, \citenamefont {Newman}, \citenamefont {Strogatz},\ and\
  \citenamefont {Watts}}]{Dun_01}%
  \BibitemOpen
  \bibfield  {author} {\bibinfo {author} {\bibfnamefont {D.~S.}\ \bibnamefont
  {Callaway}}, \bibinfo {author} {\bibfnamefont {M.~E.~J.}\ \bibnamefont
  {Newman}}, \bibinfo {author} {\bibfnamefont {S.~H.}\ \bibnamefont
  {Strogatz}}, \ and\ \bibinfo {author} {\bibfnamefont {D.~J.}\ \bibnamefont
  {Watts}},\ }\href@noop {} {\bibfield  {journal} {\bibinfo  {journal}
  {Physical Review Letters}\ }\textbf {\bibinfo {volume} {85}},\ \bibinfo
  {pages} {5468} (\bibinfo {year} {2000})}\BibitemShut {NoStop}%
\bibitem [{\citenamefont {Cohen}\ \emph {et~al.}(2002)\citenamefont {Cohen},
  \citenamefont {Havlin},\ and\ \citenamefont {ben Avraham}}]{Coh_hand}%
  \BibitemOpen
  \bibfield  {author} {\bibinfo {author} {\bibfnamefont {R.}~\bibnamefont
  {Cohen}}, \bibinfo {author} {\bibfnamefont {S.}~\bibnamefont {Havlin}}, \
  and\ \bibinfo {author} {\bibfnamefont {D.}~\bibnamefont {ben Avraham}},\
  }\enquote {\bibinfo {title} {Handbook of graphs and networks},}\ \ (\bibinfo
  {publisher} {Wiley-VCH, Berlin},\ \bibinfo {year} {2002})\ Chap.\ \bibinfo
  {chapter} {Structural properties of scale free networks}\BibitemShut
  {NoStop}%
\bibitem [{\citenamefont {Newman}\ \emph {et~al.}(2001)\citenamefont {Newman},
  \citenamefont {Strogatz},\ and\ \citenamefont {Watts}}]{New_03}%
  \BibitemOpen
  \bibfield  {author} {\bibinfo {author} {\bibfnamefont {M.~E.~J.}\
  \bibnamefont {Newman}}, \bibinfo {author} {\bibfnamefont {S.~H.}\
  \bibnamefont {Strogatz}}, \ and\ \bibinfo {author} {\bibfnamefont {D.~J.}\
  \bibnamefont {Watts}},\ }\href@noop {} {\bibfield  {journal} {\bibinfo
  {journal} {Physical Review E}\ }\textbf {\bibinfo {volume} {64}},\ \bibinfo
  {pages} {026118} (\bibinfo {year} {2001})}\BibitemShut {NoStop}%
\bibitem [{\citenamefont {Braunstein}\ \emph {et~al.}(2007)\citenamefont
  {Braunstein}, \citenamefont {Wu}, \citenamefont {Chen}, \citenamefont
  {Buldyrev}, \citenamefont {Kalisky}, \citenamefont {Sreenivasan},
  \citenamefont {Cohen}, \citenamefont {L{\'o}pez}, \citenamefont {Havlin},\
  and\ \citenamefont {Stanley}}]{Bra01}%
  \BibitemOpen
  \bibfield  {author} {\bibinfo {author} {\bibfnamefont {L.~A.}\ \bibnamefont
  {Braunstein}}, \bibinfo {author} {\bibfnamefont {Z.}~\bibnamefont {Wu}},
  \bibinfo {author} {\bibfnamefont {Y.}~\bibnamefont {Chen}}, \bibinfo {author}
  {\bibfnamefont {S.~V.}\ \bibnamefont {Buldyrev}}, \bibinfo {author}
  {\bibfnamefont {T.}~\bibnamefont {Kalisky}}, \bibinfo {author} {\bibfnamefont
  {S.}~\bibnamefont {Sreenivasan}}, \bibinfo {author} {\bibfnamefont
  {R.}~\bibnamefont {Cohen}}, \bibinfo {author} {\bibfnamefont
  {E.}~\bibnamefont {L{\'o}pez}}, \bibinfo {author} {\bibfnamefont
  {S.}~\bibnamefont {Havlin}}, \ and\ \bibinfo {author} {\bibfnamefont {H.~E.}\
  \bibnamefont {Stanley}},\ }\href@noop {} {\bibfield  {journal} {\bibinfo
  {journal} {I. J. Bifurcation and Chaos}\ }\textbf {\bibinfo {volume} {17}},\
  \bibinfo {pages} {2215} (\bibinfo {year} {2007})}\BibitemShut {NoStop}%
\bibitem [{\citenamefont {Castellano}\ and\ \citenamefont
  {Pastor-Satorras}(2010)}]{castellano_10}%
  \BibitemOpen
  \bibfield  {author} {\bibinfo {author} {\bibfnamefont {C.}~\bibnamefont
  {Castellano}}\ and\ \bibinfo {author} {\bibfnamefont {R.}~\bibnamefont
  {Pastor-Satorras}},\ }\href@noop {} {\bibfield  {journal} {\bibinfo
  {journal} {Physical review letters}\ }\textbf {\bibinfo {volume} {105}},\
  \bibinfo {pages} {218701} (\bibinfo {year} {2010})}\BibitemShut {NoStop}%
\bibitem [{\citenamefont {Pastor-Satorras}\ \emph {et~al.}(2015)\citenamefont
  {Pastor-Satorras}, \citenamefont {Castellano}, \citenamefont {Van~Mieghem},\
  and\ \citenamefont {Vespignani}}]{Pastor_15}%
  \BibitemOpen
  \bibfield  {author} {\bibinfo {author} {\bibfnamefont {R.}~\bibnamefont
  {Pastor-Satorras}}, \bibinfo {author} {\bibfnamefont {C.}~\bibnamefont
  {Castellano}}, \bibinfo {author} {\bibfnamefont {P.}~\bibnamefont
  {Van~Mieghem}}, \ and\ \bibinfo {author} {\bibfnamefont {A.}~\bibnamefont
  {Vespignani}},\ }\href@noop {} {\bibfield  {journal} {\bibinfo  {journal}
  {Rev. Mod. Phys.}\ }\textbf {\bibinfo {volume} {87}},\ \bibinfo {pages} {925}
  (\bibinfo {year} {2015})}\BibitemShut {NoStop}%
\bibitem [{\citenamefont {De~Domenico}\ \emph {et~al.}(2016)\citenamefont
  {De~Domenico}, \citenamefont {Granell}, \citenamefont {Porter},\ and\
  \citenamefont {Arenas}}]{Arenas_16}%
  \BibitemOpen
  \bibfield  {author} {\bibinfo {author} {\bibfnamefont {M.}~\bibnamefont
  {De~Domenico}}, \bibinfo {author} {\bibfnamefont {C.}~\bibnamefont
  {Granell}}, \bibinfo {author} {\bibfnamefont {M.~A.}\ \bibnamefont {Porter}},
  \ and\ \bibinfo {author} {\bibfnamefont {A.}~\bibnamefont {Arenas}},\
  }\href@noop {} {\bibfield  {journal} {\bibinfo  {journal} {Nature Physics}\
  }\textbf {\bibinfo {volume} {12}},\ \bibinfo {pages} {901} (\bibinfo {year}
  {2016})}\BibitemShut {NoStop}%
\bibitem [{\citenamefont {Wang}\ \emph {et~al.}(2017)\citenamefont {Wang},
  \citenamefont {Tang}, \citenamefont {Stanley},\ and\ \citenamefont
  {Braunstein}}]{Braunstein_16}%
  \BibitemOpen
  \bibfield  {author} {\bibinfo {author} {\bibfnamefont {W.}~\bibnamefont
  {Wang}}, \bibinfo {author} {\bibfnamefont {M.}~\bibnamefont {Tang}}, \bibinfo
  {author} {\bibfnamefont {H.~E.}\ \bibnamefont {Stanley}}, \ and\ \bibinfo
  {author} {\bibfnamefont {L.~A.}\ \bibnamefont {Braunstein}},\ }\href
  {http://stacks.iop.org/0034-4885/80/i=3/a=036603} {\bibfield  {journal}
  {\bibinfo  {journal} {Reports on Progress in Physics}\ }\textbf {\bibinfo
  {volume} {80}},\ \bibinfo {pages} {036603} (\bibinfo {year}
  {2017})}\BibitemShut {NoStop}%
\bibitem [{\citenamefont {Buono}\ \emph {et~al.}(2013)\citenamefont {Buono},
  \citenamefont {Vazquez}, \citenamefont {Macri},\ and\ \citenamefont
  {Braunstein}}]{Buo_13}%
  \BibitemOpen
  \bibfield  {author} {\bibinfo {author} {\bibfnamefont {C.}~\bibnamefont
  {Buono}}, \bibinfo {author} {\bibfnamefont {F.}~\bibnamefont {Vazquez}},
  \bibinfo {author} {\bibfnamefont {P.~A.}\ \bibnamefont {Macri}}, \ and\
  \bibinfo {author} {\bibfnamefont {L.~A.}\ \bibnamefont {Braunstein}},\
  }\href@noop {} {\bibfield  {journal} {\bibinfo  {journal} {Phys. Rev. E}\
  }\textbf {\bibinfo {volume} {88}},\ \bibinfo {pages} {022813} (\bibinfo
  {year} {2013})}\BibitemShut {NoStop}%
\bibitem [{\citenamefont {Granell}\ \emph {et~al.}(2013)\citenamefont
  {Granell}, \citenamefont {G\'omez},\ and\ \citenamefont
  {Arenas}}]{Granell_13_1}%
  \BibitemOpen
  \bibfield  {author} {\bibinfo {author} {\bibfnamefont {C.}~\bibnamefont
  {Granell}}, \bibinfo {author} {\bibfnamefont {S.}~\bibnamefont {G\'omez}}, \
  and\ \bibinfo {author} {\bibfnamefont {A.}~\bibnamefont {Arenas}},\
  }\href@noop {} {\bibfield  {journal} {\bibinfo  {journal} {Phys. Rev. Lett.}\
  }\textbf {\bibinfo {volume} {111}},\ \bibinfo {pages} {128701} (\bibinfo
  {year} {2013})}\BibitemShut {NoStop}%
\bibitem [{\citenamefont {Cozzo}\ \emph {et~al.}(2013)\citenamefont {Cozzo},
  \citenamefont {Ba{\~n}os}, \citenamefont {Meloni},\ and\ \citenamefont
  {Moreno}}]{Cozzo_13}%
  \BibitemOpen
  \bibfield  {author} {\bibinfo {author} {\bibfnamefont {E.}~\bibnamefont
  {Cozzo}}, \bibinfo {author} {\bibfnamefont {R.~A.}\ \bibnamefont
  {Ba{\~n}os}}, \bibinfo {author} {\bibfnamefont {S.}~\bibnamefont {Meloni}}, \
  and\ \bibinfo {author} {\bibfnamefont {Y.}~\bibnamefont {Moreno}},\
  }\href@noop {} {\bibfield  {journal} {\bibinfo  {journal} {Phys. Rev. E}\
  }\textbf {\bibinfo {volume} {88}},\ \bibinfo {pages} {050801(R)} (\bibinfo
  {year} {2013})}\BibitemShut {NoStop}%
\bibitem [{\citenamefont {Sanz}\ \emph {et~al.}(2014)\citenamefont {Sanz},
  \citenamefont {Xia}, \citenamefont {Meloni},\ and\ \citenamefont
  {Moreno}}]{Sanz_14}%
  \BibitemOpen
  \bibfield  {author} {\bibinfo {author} {\bibfnamefont {J.}~\bibnamefont
  {Sanz}}, \bibinfo {author} {\bibfnamefont {C.-Y.}\ \bibnamefont {Xia}},
  \bibinfo {author} {\bibfnamefont {S.}~\bibnamefont {Meloni}}, \ and\ \bibinfo
  {author} {\bibfnamefont {Y.}~\bibnamefont {Moreno}},\ }\href@noop {}
  {\bibfield  {journal} {\bibinfo  {journal} {Physical Review X}\ }\textbf
  {\bibinfo {volume} {4}},\ \bibinfo {pages} {041005} (\bibinfo {year}
  {2014})}\BibitemShut {NoStop}%
\bibitem [{\citenamefont {Lagorio}\ \emph {et~al.}(2011)\citenamefont
  {Lagorio}, \citenamefont {Dickison}, \citenamefont {Vazquez}, \citenamefont
  {Braunstein}, \citenamefont {Macri}, \citenamefont {Migueles}, \citenamefont
  {Havlin},\ and\ \citenamefont {Stanley}}]{Lag_01}%
  \BibitemOpen
  \bibfield  {author} {\bibinfo {author} {\bibfnamefont {C.}~\bibnamefont
  {Lagorio}}, \bibinfo {author} {\bibfnamefont {M.}~\bibnamefont {Dickison}},
  \bibinfo {author} {\bibfnamefont {F.}~\bibnamefont {Vazquez}}, \bibinfo
  {author} {\bibfnamefont {L.~A.}\ \bibnamefont {Braunstein}}, \bibinfo
  {author} {\bibfnamefont {P.~A.}\ \bibnamefont {Macri}}, \bibinfo {author}
  {\bibfnamefont {M.~V.}\ \bibnamefont {Migueles}}, \bibinfo {author}
  {\bibfnamefont {S.}~\bibnamefont {Havlin}}, \ and\ \bibinfo {author}
  {\bibfnamefont {H.~E.}\ \bibnamefont {Stanley}},\ }\href@noop {} {\bibfield
  {journal} {\bibinfo  {journal} {Phys. Rev. E}\ }\textbf {\bibinfo {volume}
  {83}},\ \bibinfo {pages} {026102} (\bibinfo {year} {2011})}\BibitemShut
  {NoStop}%
\bibitem [{\citenamefont {Wang}\ \emph
  {et~al.}(2015{\natexlab{a}})\citenamefont {Wang}, \citenamefont {Andrews},
  \citenamefont {Wu}, \citenamefont {Wang},\ and\ \citenamefont
  {Bauch}}]{wang_coupled}%
  \BibitemOpen
  \bibfield  {author} {\bibinfo {author} {\bibfnamefont {Z.}~\bibnamefont
  {Wang}}, \bibinfo {author} {\bibfnamefont {M.~A.}\ \bibnamefont {Andrews}},
  \bibinfo {author} {\bibfnamefont {Z.-X.}\ \bibnamefont {Wu}}, \bibinfo
  {author} {\bibfnamefont {L.}~\bibnamefont {Wang}}, \ and\ \bibinfo {author}
  {\bibfnamefont {C.~T.}\ \bibnamefont {Bauch}},\ }\href@noop {} {\bibfield
  {journal} {\bibinfo  {journal} {Physics of life reviews}\ }\textbf {\bibinfo
  {volume} {15}},\ \bibinfo {pages} {1} (\bibinfo {year}
  {2015}{\natexlab{a}})}\BibitemShut {NoStop}%
\bibitem [{\citenamefont {Valdez}\ \emph {et~al.}(2013)\citenamefont {Valdez},
  \citenamefont {Macri},\ and\ \citenamefont {Braunstein}}]{Valdez_13}%
  \BibitemOpen
  \bibfield  {author} {\bibinfo {author} {\bibfnamefont {L.}~\bibnamefont
  {Valdez}}, \bibinfo {author} {\bibfnamefont {P.}~\bibnamefont {Macri}}, \
  and\ \bibinfo {author} {\bibfnamefont {L.}~\bibnamefont {Braunstein}},\
  }\href {\doibase https://doi.org/10.1016/j.physa.2013.05.003} {\bibfield
  {journal} {\bibinfo  {journal} {Physica A: Statistical Mechanics and its
  Applications}\ }\textbf {\bibinfo {volume} {392}},\ \bibinfo {pages} {4172 }
  (\bibinfo {year} {2013})}\BibitemShut {NoStop}%
\bibitem [{\citenamefont {Wang}\ \emph {et~al.}(2012)\citenamefont {Wang},
  \citenamefont {Zhang}, \citenamefont {Huang},\ and\ \citenamefont
  {Li}}]{wang_12}%
  \BibitemOpen
  \bibfield  {author} {\bibinfo {author} {\bibfnamefont {L.}~\bibnamefont
  {Wang}}, \bibinfo {author} {\bibfnamefont {Y.}~\bibnamefont {Zhang}},
  \bibinfo {author} {\bibfnamefont {T.}~\bibnamefont {Huang}}, \ and\ \bibinfo
  {author} {\bibfnamefont {X.}~\bibnamefont {Li}},\ }\href@noop {} {\bibfield
  {journal} {\bibinfo  {journal} {Physical Review E}\ }\textbf {\bibinfo
  {volume} {86}},\ \bibinfo {pages} {032901} (\bibinfo {year}
  {2012})}\BibitemShut {NoStop}%
\bibitem [{\citenamefont {Wang}\ \emph {et~al.}(2016)\citenamefont {Wang},
  \citenamefont {Bauch}, \citenamefont {Bhattacharyya}, \citenamefont
  {d'Onofrio}, \citenamefont {Manfredi}, \citenamefont {Perc}, \citenamefont
  {Perra}, \citenamefont {Salath{\'e}},\ and\ \citenamefont
  {Zhao}}]{vaccination16}%
  \BibitemOpen
  \bibfield  {author} {\bibinfo {author} {\bibfnamefont {Z.}~\bibnamefont
  {Wang}}, \bibinfo {author} {\bibfnamefont {C.~T.}\ \bibnamefont {Bauch}},
  \bibinfo {author} {\bibfnamefont {S.}~\bibnamefont {Bhattacharyya}}, \bibinfo
  {author} {\bibfnamefont {A.}~\bibnamefont {d'Onofrio}}, \bibinfo {author}
  {\bibfnamefont {P.}~\bibnamefont {Manfredi}}, \bibinfo {author}
  {\bibfnamefont {M.}~\bibnamefont {Perc}}, \bibinfo {author} {\bibfnamefont
  {N.}~\bibnamefont {Perra}}, \bibinfo {author} {\bibfnamefont
  {M.}~\bibnamefont {Salath{\'e}}}, \ and\ \bibinfo {author} {\bibfnamefont
  {D.}~\bibnamefont {Zhao}},\ }\href@noop {} {\bibfield  {journal} {\bibinfo
  {journal} {Physics Reports}\ }\textbf {\bibinfo {volume} {664}},\ \bibinfo
  {pages} {1} (\bibinfo {year} {2016})}\BibitemShut {NoStop}%
\bibitem [{\citenamefont {Pastor-Satorras}\ and\ \citenamefont
  {Vespignani}(2002)}]{past_03}%
  \BibitemOpen
  \bibfield  {author} {\bibinfo {author} {\bibfnamefont {R.}~\bibnamefont
  {Pastor-Satorras}}\ and\ \bibinfo {author} {\bibfnamefont {A.}~\bibnamefont
  {Vespignani}},\ }\href@noop {} {\bibfield  {journal} {\bibinfo  {journal}
  {Phys. Rev. E}\ }\textbf {\bibinfo {volume} {65}},\ \bibinfo {pages} {036104}
  (\bibinfo {year} {2002})}\BibitemShut {NoStop}%
\bibitem [{\citenamefont {Madar}\ \emph {et~al.}(2004)\citenamefont {Madar},
  \citenamefont {Kalisky}, \citenamefont {Cohen}, \citenamefont {ben Avraham},\
  and\ \citenamefont {Havlin}}]{madar_04}%
  \BibitemOpen
  \bibfield  {author} {\bibinfo {author} {\bibfnamefont {N.}~\bibnamefont
  {Madar}}, \bibinfo {author} {\bibfnamefont {T.}~\bibnamefont {Kalisky}},
  \bibinfo {author} {\bibfnamefont {R.}~\bibnamefont {Cohen}}, \bibinfo
  {author} {\bibfnamefont {D.}~\bibnamefont {ben Avraham}}, \ and\ \bibinfo
  {author} {\bibfnamefont {S.}~\bibnamefont {Havlin}},\ }\href@noop {}
  {\bibfield  {journal} {\bibinfo  {journal} {The European physical journal
  b-condensed matter and complex systems}\ }\textbf {\bibinfo {volume} {38}},\
  \bibinfo {pages} {269} (\bibinfo {year} {2004})}\BibitemShut {NoStop}%
\bibitem [{\citenamefont {Cohen}\ \emph {et~al.}(2003)\citenamefont {Cohen},
  \citenamefont {Havlin},\ and\ \citenamefont {ben Avraham}}]{Coh_03}%
  \BibitemOpen
  \bibfield  {author} {\bibinfo {author} {\bibfnamefont {R.}~\bibnamefont
  {Cohen}}, \bibinfo {author} {\bibfnamefont {S.}~\bibnamefont {Havlin}}, \
  and\ \bibinfo {author} {\bibfnamefont {D.}~\bibnamefont {ben Avraham}},\
  }\href@noop {} {\bibfield  {journal} {\bibinfo  {journal} {Phys. Rev. Lett.}\
  }\textbf {\bibinfo {volume} {91}},\ \bibinfo {pages} {247901} (\bibinfo
  {year} {2003})}\BibitemShut {NoStop}%
\bibitem [{\citenamefont {Dickison}\ \emph {et~al.}(2012)\citenamefont
  {Dickison}, \citenamefont {Havlin},\ and\ \citenamefont
  {Stanley}}]{Dickison_12}%
  \BibitemOpen
  \bibfield  {author} {\bibinfo {author} {\bibfnamefont {M.}~\bibnamefont
  {Dickison}}, \bibinfo {author} {\bibfnamefont {S.}~\bibnamefont {Havlin}}, \
  and\ \bibinfo {author} {\bibfnamefont {H.~E.}\ \bibnamefont {Stanley}},\
  }\href@noop {} {\bibfield  {journal} {\bibinfo  {journal} {Physical Review
  E}\ }\textbf {\bibinfo {volume} {85}},\ \bibinfo {pages} {066109} (\bibinfo
  {year} {2012})}\BibitemShut {NoStop}%
\bibitem [{\citenamefont {Saumell-Mendiola}\ \emph {et~al.}(2012)\citenamefont
  {Saumell-Mendiola}, \citenamefont {Serrano},\ and\ \citenamefont
  {Bogu{\~n}{\'a}}}]{Men_12}%
  \BibitemOpen
  \bibfield  {author} {\bibinfo {author} {\bibfnamefont {A.}~\bibnamefont
  {Saumell-Mendiola}}, \bibinfo {author} {\bibfnamefont {M.~{\'A}.}\
  \bibnamefont {Serrano}}, \ and\ \bibinfo {author} {\bibfnamefont
  {M.}~\bibnamefont {Bogu{\~n}{\'a}}},\ }\href@noop {} {\bibfield  {journal}
  {\bibinfo  {journal} {Physical Review E}\ }\textbf {\bibinfo {volume} {86}},\
  \bibinfo {pages} {026106} (\bibinfo {year} {2012})}\BibitemShut {NoStop}%
\bibitem [{\citenamefont {Yagan}\ \emph {et~al.}(2013)\citenamefont {Yagan},
  \citenamefont {Qian}, \citenamefont {Zhang},\ and\ \citenamefont
  {Cochran}}]{Yag_13}%
  \BibitemOpen
  \bibfield  {author} {\bibinfo {author} {\bibfnamefont {O.}~\bibnamefont
  {Yagan}}, \bibinfo {author} {\bibfnamefont {D.}~\bibnamefont {Qian}},
  \bibinfo {author} {\bibfnamefont {J.}~\bibnamefont {Zhang}}, \ and\ \bibinfo
  {author} {\bibfnamefont {D.}~\bibnamefont {Cochran}},\ }\href@noop {}
  {\bibfield  {journal} {\bibinfo  {journal} {IEEE JSAC Special Issue on
  Network Science}\ }\textbf {\bibinfo {volume} {31}},\ \bibinfo {pages} {1038}
  (\bibinfo {year} {2013})}\BibitemShut {NoStop}%
\bibitem [{\citenamefont {Buono}\ \emph {et~al.}(2014)\citenamefont {Buono},
  \citenamefont {Alvarez-Zuzek}, \citenamefont {Braunstein},\ and\
  \citenamefont {Macri}}]{Buono_14}%
  \BibitemOpen
  \bibfield  {author} {\bibinfo {author} {\bibfnamefont {C.}~\bibnamefont
  {Buono}}, \bibinfo {author} {\bibfnamefont {L.~G.}\ \bibnamefont
  {Alvarez-Zuzek}}, \bibinfo {author} {\bibfnamefont {L.~A.}\ \bibnamefont
  {Braunstein}}, \ and\ \bibinfo {author} {\bibfnamefont {P.~A.}\ \bibnamefont
  {Macri}},\ }\href@noop {} {\bibfield  {journal} {\bibinfo  {journal} {PLOS
  ONE}\ }\textbf {\bibinfo {volume} {9}},\ \bibinfo {pages} {e92200} (\bibinfo
  {year} {2014})}\BibitemShut {NoStop}%
\bibitem [{\citenamefont {de~Arruda}\ \emph {et~al.}(2017)\citenamefont
  {de~Arruda}, \citenamefont {Cozzo}, \citenamefont {Peixoto}, \citenamefont
  {Rodrigues},\ and\ \citenamefont {Moreno}}]{arruda_17}%
  \BibitemOpen
  \bibfield  {author} {\bibinfo {author} {\bibfnamefont {G.~F.}\ \bibnamefont
  {de~Arruda}}, \bibinfo {author} {\bibfnamefont {E.}~\bibnamefont {Cozzo}},
  \bibinfo {author} {\bibfnamefont {T.~P.}\ \bibnamefont {Peixoto}}, \bibinfo
  {author} {\bibfnamefont {F.~A.}\ \bibnamefont {Rodrigues}}, \ and\ \bibinfo
  {author} {\bibfnamefont {Y.}~\bibnamefont {Moreno}},\ }\href@noop {}
  {\bibfield  {journal} {\bibinfo  {journal} {Physical Review X}\ }\textbf
  {\bibinfo {volume} {7}},\ \bibinfo {pages} {011014} (\bibinfo {year}
  {2017})}\BibitemShut {NoStop}%
\bibitem [{\citenamefont {Tizzoni}\ \emph {et~al.}(2012)\citenamefont
  {Tizzoni}, \citenamefont {Bajardi}, \citenamefont {Poletto}, \citenamefont
  {Ramasco}, \citenamefont {Balcan}, \citenamefont {Gon{\c{c}}alves},
  \citenamefont {Perra}, \citenamefont {Colizza},\ and\ \citenamefont
  {Vespignani}}]{tizzoni2012real}%
  \BibitemOpen
  \bibfield  {author} {\bibinfo {author} {\bibfnamefont {M.}~\bibnamefont
  {Tizzoni}}, \bibinfo {author} {\bibfnamefont {P.}~\bibnamefont {Bajardi}},
  \bibinfo {author} {\bibfnamefont {C.}~\bibnamefont {Poletto}}, \bibinfo
  {author} {\bibfnamefont {J.~J.}\ \bibnamefont {Ramasco}}, \bibinfo {author}
  {\bibfnamefont {D.}~\bibnamefont {Balcan}}, \bibinfo {author} {\bibfnamefont
  {B.}~\bibnamefont {Gon{\c{c}}alves}}, \bibinfo {author} {\bibfnamefont
  {N.}~\bibnamefont {Perra}}, \bibinfo {author} {\bibfnamefont
  {V.}~\bibnamefont {Colizza}}, \ and\ \bibinfo {author} {\bibfnamefont
  {A.}~\bibnamefont {Vespignani}},\ }\href@noop {} {\bibfield  {journal}
  {\bibinfo  {journal} {BMC medicine}\ }\textbf {\bibinfo {volume} {10}},\
  \bibinfo {pages} {165} (\bibinfo {year} {2012})}\BibitemShut {NoStop}%
\bibitem [{\citenamefont {Valdez}\ \emph {et~al.}(2014)\citenamefont {Valdez},
  \citenamefont {Arag{\~a}o}, \citenamefont {Stanley},\ and\ \citenamefont
  {Braunstein}}]{Valdez_Ebola}%
  \BibitemOpen
  \bibfield  {author} {\bibinfo {author} {\bibfnamefont {L.}~\bibnamefont
  {Valdez}}, \bibinfo {author} {\bibfnamefont {R.~H.}\ \bibnamefont
  {Arag{\~a}o}}, \bibinfo {author} {\bibfnamefont {H.}~\bibnamefont {Stanley}},
  \ and\ \bibinfo {author} {\bibfnamefont {L.}~\bibnamefont {Braunstein}},\
  }\href@noop {} {\bibfield  {journal} {\bibinfo  {journal} {Scientific
  reports}\ }\textbf {\bibinfo {volume} {5}},\ \bibinfo {pages} {12172}
  (\bibinfo {year} {2014})}\BibitemShut {NoStop}%
\bibitem [{\citenamefont {Gomes}\ \emph {et~al.}(2014)\citenamefont {Gomes},
  \citenamefont {y~Piontti}, \citenamefont {Rossi}, \citenamefont {Chao},
  \citenamefont {Longini}, \citenamefont {Halloran},\ and\ \citenamefont
  {Vespignani}}]{Gomes_14}%
  \BibitemOpen
  \bibfield  {author} {\bibinfo {author} {\bibfnamefont {M.~F.~C.}\
  \bibnamefont {Gomes}}, \bibinfo {author} {\bibfnamefont {A.~P.}\ \bibnamefont
  {y~Piontti}}, \bibinfo {author} {\bibfnamefont {L.}~\bibnamefont {Rossi}},
  \bibinfo {author} {\bibfnamefont {D.}~\bibnamefont {Chao}}, \bibinfo {author}
  {\bibfnamefont {I.}~\bibnamefont {Longini}}, \bibinfo {author} {\bibfnamefont
  {M.~E.}\ \bibnamefont {Halloran}}, \ and\ \bibinfo {author} {\bibfnamefont
  {A.}~\bibnamefont {Vespignani}},\ }\href@noop {} {\bibfield  {journal}
  {\bibinfo  {journal} {PLOS Current Outbreaks}\ } (\bibinfo {year} {2014})},\
  \bibinfo {note}
  {doi:10.1371/currents.outbreaks.cd818f63d40e24aef769dda7df9e0da5}\BibitemShut
  {NoStop}%
\bibitem [{\citenamefont {Alvarez-Zuzek}\ \emph
  {et~al.}(2015{\natexlab{a}})\citenamefont {Alvarez-Zuzek}, \citenamefont
  {Stanley},\ and\ \citenamefont {Braunstein}}]{Alvarez-zuzek_15}%
  \BibitemOpen
  \bibfield  {author} {\bibinfo {author} {\bibfnamefont {L.~G.}\ \bibnamefont
  {Alvarez-Zuzek}}, \bibinfo {author} {\bibfnamefont {H.~E.}\ \bibnamefont
  {Stanley}}, \ and\ \bibinfo {author} {\bibfnamefont {L.~A.}\ \bibnamefont
  {Braunstein}},\ }\href@noop {} {\bibfield  {journal} {\bibinfo  {journal}
  {Scientific Reports}\ }\textbf {\bibinfo {volume} {5}},\ \bibinfo {pages}
  {12151} (\bibinfo {year} {2015}{\natexlab{a}})}\BibitemShut {NoStop}%
\bibitem [{\citenamefont {Wang}\ \emph
  {et~al.}(2015{\natexlab{b}})\citenamefont {Wang}, \citenamefont {Zhao},
  \citenamefont {Wang}, \citenamefont {Sun},\ and\ \citenamefont
  {Jin}}]{wang_acquaintance}%
  \BibitemOpen
  \bibfield  {author} {\bibinfo {author} {\bibfnamefont {Z.}~\bibnamefont
  {Wang}}, \bibinfo {author} {\bibfnamefont {D.-W.}\ \bibnamefont {Zhao}},
  \bibinfo {author} {\bibfnamefont {L.}~\bibnamefont {Wang}}, \bibinfo {author}
  {\bibfnamefont {G.-Q.}\ \bibnamefont {Sun}}, \ and\ \bibinfo {author}
  {\bibfnamefont {Z.}~\bibnamefont {Jin}},\ }\href@noop {} {\bibfield
  {journal} {\bibinfo  {journal} {EPL (Europhysics Letters)}\ }\textbf
  {\bibinfo {volume} {112}},\ \bibinfo {pages} {48002} (\bibinfo {year}
  {2015}{\natexlab{b}})}\BibitemShut {NoStop}%
\bibitem [{\citenamefont {Fenner}\ \emph {et~al.}(1988)\citenamefont {Fenner},
  \citenamefont {Henderson}, \citenamefont {Arita}, \citenamefont {Jezek},\
  and\ \citenamefont {Ladnyi}}]{Smallpox}%
  \BibitemOpen
  \bibfield  {author} {\bibinfo {author} {\bibfnamefont {F.}~\bibnamefont
  {Fenner}}, \bibinfo {author} {\bibfnamefont {D.~A.}\ \bibnamefont
  {Henderson}}, \bibinfo {author} {\bibfnamefont {I.}~\bibnamefont {Arita}},
  \bibinfo {author} {\bibfnamefont {Z.}~\bibnamefont {Jezek}}, \ and\ \bibinfo
  {author} {\bibfnamefont {I.~D.}\ \bibnamefont {Ladnyi}},\ }\href@noop {}
  {\emph {\bibinfo {title} {Smallpox and its eradication}}}\ (\bibinfo
  {publisher} {World Health Organization},\ \bibinfo {year} {1988})\BibitemShut
  {NoStop}%
\bibitem [{\citenamefont {Perisic}\ and\ \citenamefont
  {Bauch}(2009)}]{perisic_09}%
  \BibitemOpen
  \bibfield  {author} {\bibinfo {author} {\bibfnamefont {A.}~\bibnamefont
  {Perisic}}\ and\ \bibinfo {author} {\bibfnamefont {C.~T.}\ \bibnamefont
  {Bauch}},\ }\href@noop {} {\bibfield  {journal} {\bibinfo  {journal} {PLoS
  computational biology}\ }\textbf {\bibinfo {volume} {5}},\ \bibinfo {pages}
  {e1000280} (\bibinfo {year} {2009})}\BibitemShut {NoStop}%
\bibitem [{\citenamefont {M{\"u}ller}\ \emph {et~al.}(2000)\citenamefont
  {M{\"u}ller}, \citenamefont {Sch{\"o}nfisch},\ and\ \citenamefont
  {Kirkilionis}}]{muller_ring}%
  \BibitemOpen
  \bibfield  {author} {\bibinfo {author} {\bibfnamefont {J.}~\bibnamefont
  {M{\"u}ller}}, \bibinfo {author} {\bibfnamefont {B.}~\bibnamefont
  {Sch{\"o}nfisch}}, \ and\ \bibinfo {author} {\bibfnamefont {M.}~\bibnamefont
  {Kirkilionis}},\ }\href@noop {} {\bibfield  {journal} {\bibinfo  {journal}
  {Journal of mathematical biology}\ }\textbf {\bibinfo {volume} {41}},\
  \bibinfo {pages} {143} (\bibinfo {year} {2000})}\BibitemShut {NoStop}%
\bibitem [{\citenamefont {Merler}\ \emph {et~al.}(2016)\citenamefont {Merler},
  \citenamefont {Ajelli}, \citenamefont {Fumanelli}, \citenamefont
  {Parlamento}, \citenamefont {y~Piontti}, \citenamefont {Dean}, \citenamefont
  {Putoto}, \citenamefont {Carraro}, \citenamefont {Longini~Jr}, \citenamefont
  {Halloran} \emph {et~al.}}]{merler_16}%
  \BibitemOpen
  \bibfield  {author} {\bibinfo {author} {\bibfnamefont {S.}~\bibnamefont
  {Merler}}, \bibinfo {author} {\bibfnamefont {M.}~\bibnamefont {Ajelli}},
  \bibinfo {author} {\bibfnamefont {L.}~\bibnamefont {Fumanelli}}, \bibinfo
  {author} {\bibfnamefont {S.}~\bibnamefont {Parlamento}}, \bibinfo {author}
  {\bibfnamefont {A.~P.}\ \bibnamefont {y~Piontti}}, \bibinfo {author}
  {\bibfnamefont {N.~E.}\ \bibnamefont {Dean}}, \bibinfo {author}
  {\bibfnamefont {G.}~\bibnamefont {Putoto}}, \bibinfo {author} {\bibfnamefont
  {D.}~\bibnamefont {Carraro}}, \bibinfo {author} {\bibfnamefont {I.~M.}\
  \bibnamefont {Longini~Jr}}, \bibinfo {author} {\bibfnamefont {M.~E.}\
  \bibnamefont {Halloran}},  \emph {et~al.},\ }\href@noop {} {\bibfield
  {journal} {\bibinfo  {journal} {PLoS neglected tropical diseases}\ }\textbf
  {\bibinfo {volume} {10}},\ \bibinfo {pages} {e0005093} (\bibinfo {year}
  {2016})}\BibitemShut {NoStop}%
\bibitem [{\citenamefont {Gsell}\ \emph {et~al.}(2017)\citenamefont {Gsell},
  \citenamefont {Camacho}, \citenamefont {J~Kucharski}, \citenamefont
  {H~Watson}, \citenamefont {Bagayoko}, \citenamefont {Danmadji~Nadlaou} \emph
  {et~al.}}]{lancet_ebola}%
  \BibitemOpen
  \bibfield  {author} {\bibinfo {author} {\bibfnamefont {P.-S.}\ \bibnamefont
  {Gsell}}, \bibinfo {author} {\bibfnamefont {A.}~\bibnamefont {Camacho}},
  \bibinfo {author} {\bibfnamefont {A.}~\bibnamefont {J~Kucharski}}, \bibinfo
  {author} {\bibfnamefont {C.}~\bibnamefont {H~Watson}}, \bibinfo {author}
  {\bibfnamefont {A.}~\bibnamefont {Bagayoko}}, \bibinfo {author}
  {\bibfnamefont {S.}~\bibnamefont {Danmadji~Nadlaou}},  \emph {et~al.},\
  }\href@noop {} {\bibfield  {journal} {\bibinfo  {journal} {The Lancet
  Infectious Diseases}\ }\textbf {\bibinfo {volume} {17}},\ \bibinfo {pages}
  {1276} (\bibinfo {year} {2017})}\BibitemShut {NoStop}%
\bibitem [{\citenamefont {Molloy}\ and\ \citenamefont {Reed}(1995)}]{Mol_01}%
  \BibitemOpen
  \bibfield  {author} {\bibinfo {author} {\bibfnamefont {M.}~\bibnamefont
  {Molloy}}\ and\ \bibinfo {author} {\bibfnamefont {B.}~\bibnamefont {Reed}},\
  }\href@noop {} {\bibfield  {journal} {\bibinfo  {journal} {Random Structures
  and Algorithms}\ }\textbf {\bibinfo {volume} {6}},\ \bibinfo {pages} {161}
  (\bibinfo {year} {1995})}\BibitemShut {NoStop}%
\bibitem [{\citenamefont {Lagorio}\ \emph {et~al.}(2009)\citenamefont
  {Lagorio}, \citenamefont {Migueles}, \citenamefont {Braunstein},
  \citenamefont {L{\'o}pez},\ and\ \citenamefont {Macri}}]{Lag_02}%
  \BibitemOpen
  \bibfield  {author} {\bibinfo {author} {\bibfnamefont {C.}~\bibnamefont
  {Lagorio}}, \bibinfo {author} {\bibfnamefont {M.~V.}\ \bibnamefont
  {Migueles}}, \bibinfo {author} {\bibfnamefont {L.~A.}\ \bibnamefont
  {Braunstein}}, \bibinfo {author} {\bibfnamefont {E.}~\bibnamefont
  {L{\'o}pez}}, \ and\ \bibinfo {author} {\bibfnamefont {P.~A.}\ \bibnamefont
  {Macri}},\ }\href@noop {} {\bibfield  {journal} {\bibinfo  {journal} {Physica
  A}\ }\textbf {\bibinfo {volume} {388}},\ \bibinfo {pages} {755} (\bibinfo
  {year} {2009})}\BibitemShut {NoStop}%
\bibitem [{\citenamefont {Hu}\ \emph {et~al.}(2015)\citenamefont {Hu},
  \citenamefont {Ji}, \citenamefont {Feng}, \citenamefont {Havlin},\ and\
  \citenamefont {Jin}}]{Yanqing15}%
  \BibitemOpen
  \bibfield  {author} {\bibinfo {author} {\bibfnamefont {Y.}~\bibnamefont
  {Hu}}, \bibinfo {author} {\bibfnamefont {S.}~\bibnamefont {Ji}}, \bibinfo
  {author} {\bibfnamefont {L.}~\bibnamefont {Feng}}, \bibinfo {author}
  {\bibfnamefont {S.}~\bibnamefont {Havlin}}, \ and\ \bibinfo {author}
  {\bibfnamefont {Y.}~\bibnamefont {Jin}},\ }\href@noop {} {\bibfield
  {journal} {\bibinfo  {journal} {arXiv preprint arXiv:1509.03484}\ } (\bibinfo
  {year} {2015})}\BibitemShut {NoStop}%
\bibitem [{\citenamefont {Erd\H{o}s}\ and\ \citenamefont
  {R{\'e}nyi}(1959)}]{Erd_01}%
  \BibitemOpen
  \bibfield  {author} {\bibinfo {author} {\bibfnamefont {P.}~\bibnamefont
  {Erd\H{o}s}}\ and\ \bibinfo {author} {\bibfnamefont {A.}~\bibnamefont
  {R{\'e}nyi}},\ }\href@noop {} {\bibfield  {journal} {\bibinfo  {journal}
  {Publications Mathematicae}\ }\textbf {\bibinfo {volume} {6}},\ \bibinfo
  {pages} {290} (\bibinfo {year} {1959})}\BibitemShut {NoStop}%
\bibitem [{\citenamefont {Amaral}\ \emph {et~al.}(2000)\citenamefont {Amaral},
  \citenamefont {Scala}, \citenamefont {Barth{\'e}lemy},\ and\ \citenamefont
  {Stanley}}]{Ama_01}%
  \BibitemOpen
  \bibfield  {author} {\bibinfo {author} {\bibfnamefont {L.~A.~N.}\
  \bibnamefont {Amaral}}, \bibinfo {author} {\bibfnamefont {A.}~\bibnamefont
  {Scala}}, \bibinfo {author} {\bibfnamefont {M.}~\bibnamefont
  {Barth{\'e}lemy}}, \ and\ \bibinfo {author} {\bibfnamefont {H.~E.}\
  \bibnamefont {Stanley}},\ }\href@noop {} {\bibfield  {journal} {\bibinfo
  {journal} {Proc. Natl. Acad. Sci. USA}\ }\textbf {\bibinfo {volume} {97}},\
  \bibinfo {pages} {11149} (\bibinfo {year} {2000})}\BibitemShut {NoStop}%
\bibitem [{\citenamefont {Cohen}\ and\ \citenamefont
  {Havlin}(2003)}]{cohen_ultrasmall}%
  \BibitemOpen
  \bibfield  {author} {\bibinfo {author} {\bibfnamefont {R.}~\bibnamefont
  {Cohen}}\ and\ \bibinfo {author} {\bibfnamefont {S.}~\bibnamefont {Havlin}},\
  }\href@noop {} {\bibfield  {journal} {\bibinfo  {journal} {Physical review
  letters}\ }\textbf {\bibinfo {volume} {90}},\ \bibinfo {pages} {058701}
  (\bibinfo {year} {2003})}\BibitemShut {NoStop}%
\bibitem [{\citenamefont {Buono}\ and\ \citenamefont
  {Braunstein}(2015)}]{Buono_15}%
  \BibitemOpen
  \bibfield  {author} {\bibinfo {author} {\bibfnamefont {C.}~\bibnamefont
  {Buono}}\ and\ \bibinfo {author} {\bibfnamefont {L.~A.}\ \bibnamefont
  {Braunstein}},\ }\href@noop {} {\bibfield  {journal} {\bibinfo  {journal}
  {EPL (Europhysics Letters)}\ }\textbf {\bibinfo {volume} {109}},\ \bibinfo
  {pages} {26001} (\bibinfo {year} {2015})}\BibitemShut {NoStop}%
\bibitem [{\citenamefont {Alvarez-Zuzek}\ \emph
  {et~al.}(2015{\natexlab{b}})\citenamefont {Alvarez-Zuzek}, \citenamefont
  {Buono},\ and\ \citenamefont {Braunstein}}]{alvarez2015}%
  \BibitemOpen
  \bibfield  {author} {\bibinfo {author} {\bibfnamefont {L.~G.}\ \bibnamefont
  {Alvarez-Zuzek}}, \bibinfo {author} {\bibfnamefont {C.}~\bibnamefont
  {Buono}}, \ and\ \bibinfo {author} {\bibfnamefont {L.~A.}\ \bibnamefont
  {Braunstein}},\ }in\ \href@noop {} {\emph {\bibinfo {booktitle} {Journal of
  Physics Conference Series}}},\ Vol.\ \bibinfo {volume} {640}\ (\bibinfo
  {year} {2015})\BibitemShut {NoStop}%
\end{thebibliography}

%

\end{document}